\documentclass[10pt,journal,compsoc]{IEEEtran}

\usepackage[export]{adjustbox}

\ifCLASSOPTIONcompsoc
  \usepackage[nocompress]{cite}
\else
  \usepackage{cite}
\fi

\usepackage{lipsum}
\usepackage{comment}

\usepackage{amssymb,amsmath}
\usepackage{stix}

\usepackage{overpic}
\usepackage[dvipsnames]{xcolor}
\usepackage{tikz,pgfplots}
\pgfplotsset{compat=1.17}
\usetikzlibrary{arrows, arrows.meta}

\usepackage{multirow,makecell,colortbl,arydshln}

\definecolor{c1}{HTML}{c1272d}
\definecolor{c2}{HTML}{0000a7}
\definecolor{c3}{HTML}{eecc16}
\definecolor{c4}{HTML}{27E1C1}

\definecolor{americanrose}{rgb}{1.0, 0.01, 0.24}
\definecolor{mathblue}{HTML}{DC0000}
\definecolor{myred}{rgb}{0.753, 0.314, 0.275}
\definecolor{myblue}{rgb}{0.0, 0.24, 0.95}
\definecolor{tbl_gray}{gray}{0.85}

\newcommand\MYhyperrefoptions{bookmarks=true,bookmarksnumbered=true,
pdfpagemode={UseOutlines},plainpages=false,pdfpagelabels=true,
colorlinks=true,linkcolor={americanrose},citecolor={myblue},urlcolor={myblue},
pdftitle={Deep Hough Transform for Semantic Line Detection},
pdfsubject={Typesetting},
pdfauthor={Kai Zhao et al.}}
\usepackage[\MYhyperrefoptions,pdftex]{hyperref}

\usepackage{cleveref}
\crefname{equation}{Eq.}{Eq.}
\crefname{figure}{Fig.}{Fig.}
\crefname{table}{Tab.}{Tab.}
\crefname{section}{Sec.}{Sec.}

\newcommand{\lr}{low-resolution}
\newcommand{\hr}{high-resolution}
\newcommand{\sr}{super-resolution}
\newcommand{\imsr}{image~\sr}
\newcommand{\lrim}{\lr{} image}
\newcommand{\hrim}{\hr{} image}

\newcommand{\lrhrim}{low- and high-resolution images}

\newcommand{\mypar}[1]{\vspace{0.5em}\noindent\textbf{#1}.\hspace{0.2em}}

\newcommand{\stdgs}{\mathcal{N}(0, \mathbf{I})}

\begin{document}
\title{PartDiff: Image Super-resolution with Partial Diffusion Models}

\author{
    Kai Zhao,
    \and Alex Ling Yu Hung,
    \and Kaifeng Pang,
    \and Haoxin Zheng,
    \and Kyunghyun Sung
    \thanks{
        Submitted on \today{} for review.
        This work was supported in part
        by the National Institutes of Health R01-CA248506 and funds
        from the Integrated Diagnostics Program, Departments of Radiological
        Sciences and Pathology, David Geffen School of Medicine, UCLA.
    }\thanks{
        Kai Zhao and Kyunghyun Sung are with the
        Department of Radiological Sciences, University of California, Los Angeles
        (UCLA), CA 90095 USA.
        e-mail: \url{kz@kaizhao.net}, \url{ksung@mednet.ucla.edu},
        .
    }
    \thanks{
        Alex Ling Yu Hung, Kaifeng Pang, and Haoxin Zheng  are with the Departments
        of Radiological Sciences and Computer Science,
        University of California, Los Angeles (UCLA), CA 90095, USA.
        email: \{alexhung96@, calvinpang777@g., hzheng@mednet.\}ucla.edu
    }
}

\markboth{}%
{Shell \MakeLowercase{\textit{et al.}}: Bare Demo of IEEEtran.cls for Computer Society Journals}

\IEEEtitleabstractindextext{%
\begin{abstract}
    Denoising diffusion probabilistic models (DDPMs) have achieved impressive performance on various image generation tasks,
    including \imsr.
    By learning to reverse the process of gradually diffusing the data distribution into Gaussian noise,
    DDPMs generate new data by iteratively denoising from random noise.
    Despite their impressive performance, diffusion-based generative models suffer from high computational costs due to the large number of denoising steps.
    In this paper, we first observed that the intermediate latent states gradually converge and become indistinguishable when diffusing a pair of
    low- and high-resolution images.
    This observation inspired us to propose the Partial Diffusion Model (PartDiff),
    which diffuses the image to an intermediate latent state instead of pure random noise,
    where the intermediate latent state is approximated by the latent of diffusing the low-resolution image.
    During generation, Partial Diffusion Models start denoising from the intermediate distribution and perform only a part of the denoising steps.
    Additionally, to mitigate the error caused by the approximation, we introduce ‘latent alignment,’ which aligns the latent between
    low- and high-resolution images during training.
    Experiments on both magnetic resonance imaging (MRI) and natural images show that,
    compared to plain diffusion-based super-resolution methods, Partial Diffusion Models significantly reduce the number of denoising steps
    without sacrificing the quality of generation.
\end{abstract}

}
\maketitle

\IEEEpeerreviewmaketitle


\IEEEraisesectionheading{\section{Introduction}\label{sec:intro}}
Single \imsr~\cite{irani1991improving}, which aims to generate
a high-resolution (HR) raster image consistent with a low-resolution (LR) input,
is an important problem in computer vision and image processing.
However, it is an ill-posed challenging task
because a specific low-resolution input may correspond to multiple high-resolution outputs.
Many recent studies employ powerful deep neural networks (DNNs)
for \imsr{} and promising performance has been achieved.

Early DNN-based \sr{} methods train feed-forward DNNs to learn the mapping between low-
and high-resolution images~\cite{dong2015image,johnson2016perceptual} from substantial training pairs.
Although feed-forward networks can achieve impressive results at low upsampling factors, they cannot reproduce high-fidelity
details at high upsampling factors because of the highly complex distribution of output images
condition on the input images.
\begin{figure}
    \vspace{1em}
    \begin{overpic}[width=0.97\linewidth, right]{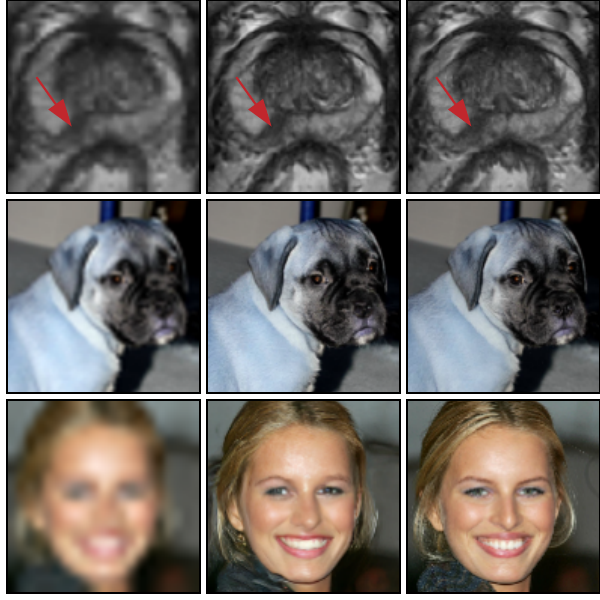}
        \put(12,98){LR Input}
        \put(43,98){SR Output}
        \put(73,98){HR Reference}
        \put(0,80){\rotatebox{90}{$\times2$}}
        \put(0,50){\rotatebox{90}{$\times4$}}
        \put(0,15){\rotatebox{90}{$\times8$}}
    \end{overpic}
    \vspace{-2.5em}
    \caption{
        $\times2, \times4$, and $\times8$ \sr{} results of T2-weighted prostate
        MRI (top), natural (middle)
        and facial images (bottom).
        The prostate cancer lesion is marked with the red arrow.
    }
    \label{fig:sr-t2-im-face}
    \vspace{-1em}
\end{figure}
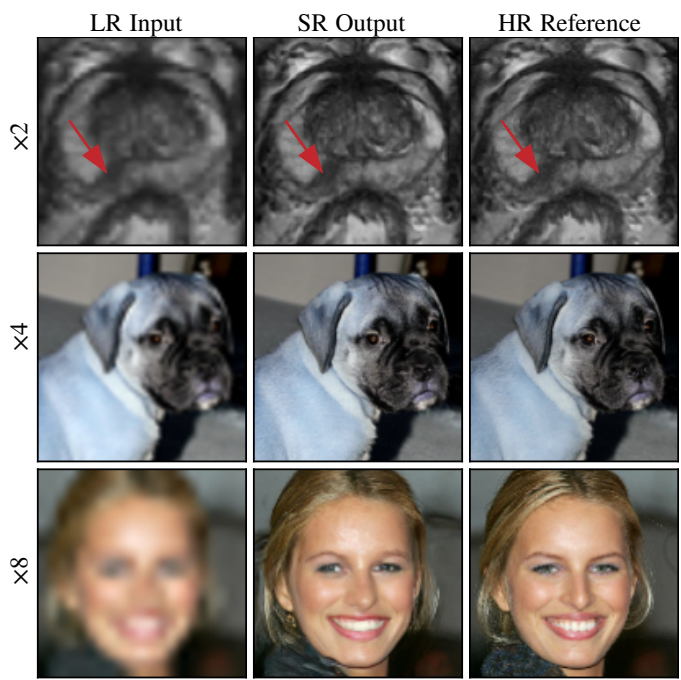
Deep generative models, including
Generative Adversarial Networks (GANs)~\cite{goodfellow2014generative}
and Variational Auto-encoders (VAEs)~\cite{kingma2013auto}
have shown impressive results in data generation of various modalities
including image~\cite{goodfellow2014generative,karras2018progressive,brock2018large}
video~\cite{kalchbrenner2017video,brooks2022generating},
and audio~\cite{vandenoord16_ssw,prenger2019waveglow}.
Generative Adversarial Networks (GANs)~\cite{goodfellow2014generative} have shown impressive results in image generation
and have been applied to various conditional image generation tasks including \imsr~\cite{chen2018fsrnet,dahl2017pixel}.
Though generating striking images, GANs often suffer from instability in model optimization and model collapse.

\begin{figure*}[!htb]
    \newcommand{\diff}[1]{{\color{c2}{\large{#1}}}}
    \newcommand{\pdiff}[1]{{\color{c1}{\large{#1}}}}
    \centering
    \vspace{2em}
    \begin{overpic}[width=0.96\linewidth,right]{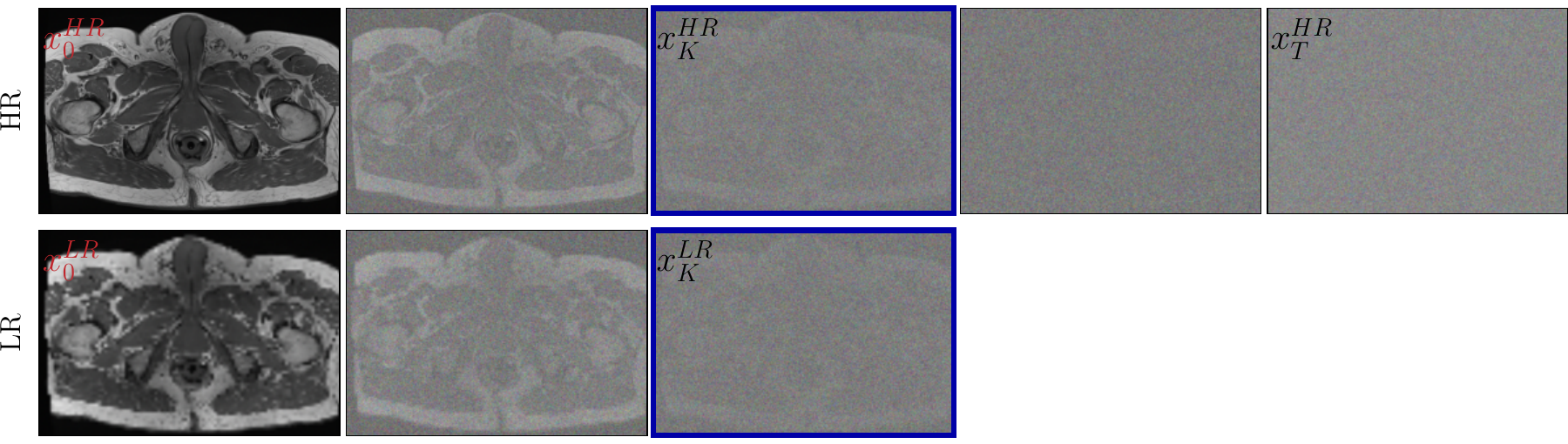}
        \put(23.8,21){\diff{$\Longleftarrow$}}
        \put(23.8,17.5){\pdiff{$\Longleftarrow$}}
        \put(42.6,21){\diff{$\Longleftarrow$}}
        \put(42.6,17.5){\pdiff{$\Longleftarrow$}}
        \put(61.0,21){\diff{$\Longleftarrow$}}
        \put(80.0,21){\diff{$\Longleftarrow$}}
        \put(22.8,6){\pdiff{$\Longrightarrow$}}
        \put(41.8,6){\pdiff{$\Longrightarrow$}}
        \put(52,12.2){\pdiff{\rotatebox{90}{$\Longrightarrow$}}}
        \put(75,7.5){\diff{$\Longrightarrow$} Diffusion}
        \put(75,4.5){\pdiff{$\Longrightarrow$} Partial Diffusion}
    \end{overpic}    
    \vspace{-25pt}
    \caption{
        The diffusion processes of high-resolution (top) and low-resolution (bottom) MRI images.
        The latent states gradually converge and become indistinguishable (images with blue frames).
        Partial diffusion models use the low-resolution latent state $x_K{LR}$
        as the proxy for high-resolution latent $x_K{HR}$
        to reduce the number of denoising steps.
    }
    \label{fig:diff-lr-hr}
    \vspace{-1em}
\end{figure*}
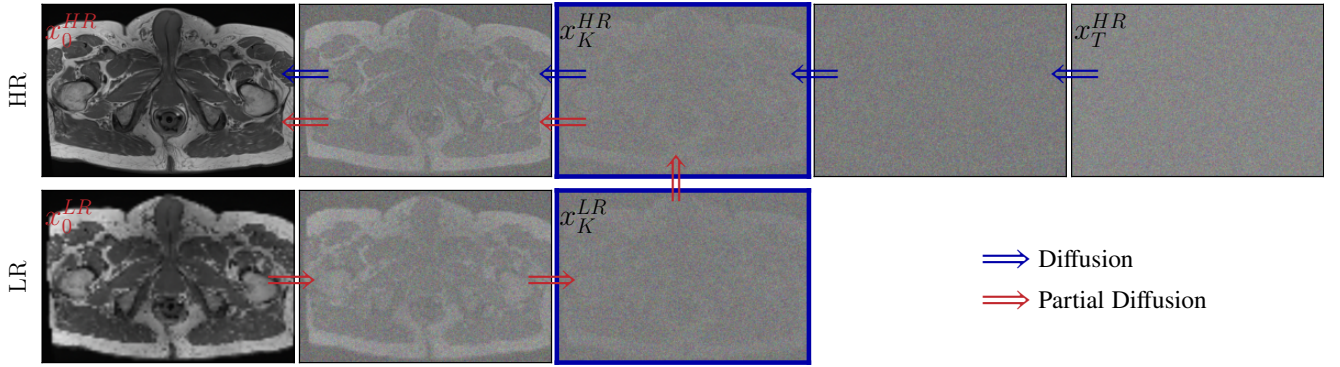

Recently, Denoising Diffusion Probabilistic Models
(DDPMs)~\cite{ho2020denoising} have shown striking
performance on image generation tasks
and have been applied to \imsr~\cite{sr3,li2022srdiff}.
By modeling the reverse process of gradually diffusing the data distribution
into Gaussian noise, diffusion models generate new data by iteratively
denoising from a random Gaussian noise.
SR3~\cite{sr3} and SRDiff~\cite{li2022srdiff} adopt diffusion models for 
\imsr{} and have shown their outstanding performance in generating
realistic high-resolution images.
One notorious deficiency of diffusion models is the tedious generation process which includes
thousands of denoising steps.
SRDiff~\cite{li2022srdiff} proposes to denoise in a low-dimensional
embedding space, but it still requires a large number of denoising steps.
Being slow in a generation has become the main obstacle to the practical application
of diffusion models to \sr.

Unlike unconditional image generation, where models are expected to generate
samples from pure noise,
in \imsr, 
a high-resolution output is generated from a low-resolution image input.
The low-resolution input exhibits similar structure, content,
and appearance as a high-resolution image,
except for certain high-frequency details that are missing.
This unique feature raises an interesting question:
is it necessary to denoise from a pure noise when applying diffusion models to \imsr?
To answer this question, we first investigate the diffusion processes
of low- and high-resolution image pairs.
As shown in~\cref{fig:diff-lr-hr}, we found that the intermediate latent states
of low- and high-resolution images gradually converge and become indistinguishable in the midway.
Based on the observation, we conjecture that the low-resolution latent states could be used as the proxy for
a high-resolution latent, and this motivates us to propose a novel sampling algorithm that only executes part of
the denoising steps.
As illustrated in~\cref{fig:diff-lr-hr},
our method use the intermediate latent of a \lrim{} ($x_K^{LR}$)
to substitute a latent of the high-resolution image ($x_K^{HR}$).
This allows us to bypass all denoising steps before $x_K^{HR}$
and accelerate the denoising process.

Although the two latent states are visually similar, there is still a subtle disparity in between
and this will inevitably have a detrimental effect on the quality of generation.
Therefore, we further propose the `latent alignment' that aligns the low- and high-resolution latent states to mitigate the disparity between the two.
In particular, we align the latent states by progressively
interpolating between low- and high-resolution latent states during training.
This allows a gradual shift from low-resolution to high-resolution in the latent space
and avoids the sudden disparity with the proxy.
In summary, the contributions of this paper are in three folds:
\vspace{-0.5em}
\begin{itemize}
    \item We qualitatively and quantitatively evaluate that the diffusion
                processes of low- and high-resolution images gradually converge in the midway
                and the intermediate latent states become indistinguishable.
    \item We are motivated to use the intermediate latent states
                of low-resolution images as the proxy for that of the high-resolution image.
                This allows us to accelerate diffusion models in training and testing by bypassing many denoising steps.
    \item We propose the `latent alignment' to mitigate the distributional disparity between
                low- and high-resolution latent states to further improve the quality of generation.
\end{itemize}
\vspace{-0.5em}
Experiments were performed on both magnetic resonance imaging (MRI) images and natural images to demonstrate that our method is able
to accelerate state-of-the-art diffusion-based \sr{} methods without incurring a significant loss in image quality.
~\cref{fig:sr-t2-im-face} shows some exemplar \sr{} results on different image modalities.

The rest of this paper is organized as follows:
~\cref{sec:related} summarizes the related works in \imsr{} and diffusion probabilistic models.
~\cref{sec:bg-diff} introduces the background about the denoising diffusion probabilistic models (DDPMs)~\cite{ho2020denoising}.
~\cref{sec:partdiff} elaborates on the proposed partial diffusion models (PartDiff)
and discusses the key components.
~\cref{sec:exp} presents experimental details and reports comparison results.
~\cref{sec:conclusion} makes a conclusion remark.

\section{Related Work}\label{sec:related}
Our method is inspired by recent works in generative models, especially diffusion-based models,
for \imsr.
In this section, we first briefly  introduce related works in \imsr{} with a focus on deep learning-based methods
and then cover some recent advances in diffusion models for \imsr.

\subsection{Image Super-resolution}
Image \sr{} is a one-to-many problem in which low-resolution input
may correspond to many high-resolution outputs.
Conventional methods for \imsr{} mainly seek to limit the solution space by leveraging prior information.
According to the priors, those \sr{} methods can be categorized into several types.
Edge-based methods generate edge-preserving \hrim s by learning image edge priors.
Various
edge features have been proposed, such as the depth and width of an edge~\cite{fattal2007image}
or gradient profile~\cite{sun2008image}.
Those methods generate \hrim s with high-fidelity edges but don't perform well on other high-frequency
structures such as textures.
Statistical-based methods exploit the statistical image information,
e.g. the heavy-tailed gradient distribution~\cite{shan2008fast},
the sparsity property of large gradients in generic images~\cite{kim2010single},
as prior for \imsr.
Example-based methods learn prior from exemplar images.
Those examples can be from external images
~\cite{freeman2002example,chang2004super},
the input image
~\cite{glasner2009super,freedman2011image}
or combined sources~\cite{yang2013fast}.

With the rapid development of deep learning techniques
in recent years, a line of deep learning-based \sr{}  approaches have
been invented with state-of-the-art performance on various benchmarks.
Many of the early explorations directly use convolutional neural networks (CNNs)
to regress a \hrim~\cite{dong2015image,wang2015deep,shi2016real,sajjadi2017enhancenet}.
Many new architectures~\cite{wang2015deep,shi2016real,li2019feedback}
and loss functions~\cite{sajjadi2017enhancenet,johnson2016perceptual} have been proposed to
improve the quality of \sr{}.
A recent work~\cite{chen2021learning} uses implicit representations to upsample images at an arbitrary scale.
Although being able to generate images close to the ground-truth,
regression-based methods tend to produce blurry images
that correlate poorly to human perception.
Deep generative models,
e.g., generative adversarial networks (GANs),
have shown impressive performance in generating high-fidelity realistic images
and benefited conditional tasks such as 
\imsr~\cite{ledig2017photo,brock2018large}.
A number of methods have been proposed to improve GAN-based \imsr{}
in network architecture
~\cite{ledig2017photo,brock2018large},
training strategies~\cite{ledig2017photo},
and domain-specific priors~\cite{chen2018fsrnet}.
Although GANs provide a promising direction, they generally suffer
from common failure cases of mode collapse~\cite{arjovsky2017wasserstein} and unstable training~\cite{metz2017unrolled,heusel2017gans}.

\subsection{Diffusion Probabilistic Models}
Diffusion probabilistic models~\cite{sohl2015deep} is a class of generative models
that match a data distribution by learning to reverse a gradual noising process.
Diffusion models have received growing attention in recent years due to its
promising results in generating high perceptual quality samples in various data modalities
including
image~\cite{ho2020denoising,nichol2021improved},
video~\cite{ho2022video_a},
audio~\cite{chen2021wavegrad},
and text~\cite{li2022diffusionlm}.
%
%
%
Diffusion models have also shown impressive results in condition image generation tasks
such \imsr~\cite{sr3,li2022srdiff}.
Concretely, SR3~\cite{sr3} takes the low-resolution image as an additional input
to the denoising network and sets up a conditional denoising framework.
SRDiff~\cite{li2022srdiff} also uses the \lrim{} as the condition but executes the diffusion
process in a lower-dimensional hidden space.
Both SR3 and SRDiff have shown impressive \sr{} results in restoring high-fidelity
details, especially under large upsampling factors.

While showing great potential in various generation tasks, diffusion models are notoriously
slow in inference because generating high-quality samples generally need hundreds or thousands of sequential
denoising steps~\cite{ho2020denoising,nichol2021improved}.
Many efforts have been made to accelerate diffusion models~\cite{chen2021wavegrad,nichol2021improved}.
Lu et al.~\cite{lu2022dpm} propose an exact formulation that analytically computes the linear part of
the solution to diffusion ordinary differentiable equations (ODEs).
Chen et al.~\cite{chen2021wavegrad} condition denoising models on continuous
noise scales rather than discrete denoising steps in DDPM~\cite{ho2020denoising},
such that separate noise schedules can be used in training and testing,
allowing flexible adjustment of denoising steps in inference.
This method requires carefully tuned noise schedules in testing and the resulting noise schedule
is unstable in different datasets.
Model distillation was also introduced to reduce the denoising steps of diffusion models~\cite{salimans2022progressive}.
More recently, Zheng et al.~\cite{zheng2023truncated} propose to add noise not until the data become pure random noise,
but until they reach a hidden noisy data distribution that can be confidently learned.
Consequently, few denoising steps are required to generate data from the hidden noise distribution.
However, learning this hidden noise distribution is more difficult than sampling from pure noise.

Our method is very different from all those general-purpose diffusion model acceleration methods.
It is specifically designed for the \imsr{} task where \lrim s are given as the condition.
We use the intermediate latent states of \lrim{} as a proxy to that of the
\hrim{} so that fewer denoising
steps are required to recover \hrim{} from \lrim.

\section{Background on Diffusion Models}\label{sec:bg-diff}
We introduce some background of diffusion models
and its application to \sr.
We adopt the notation in DDPMs~\cite{ho2020denoising} for both unconditional image generation
and conditional image generation for \imsr.

\subsection{Denoising Diffusion Probabilistic Models}
Diffusion models~\cite{song2019generative,ho2020denoising} transform data samples $x_0$
into Gaussian noise $x_T$ through a gradual noising process
and generate new data by learning to reverse this process.
The transition from data to noise is referred to as the forward process (or diffusion process),
and the opposite is called the reverse process (or denoising process).

\subsubsection{Forward process}
The forward process transforms data into Gaussian noise
by iteratively adding Gaussian noise to a clean sample $x_0$.
This can be formulated as a Markov process with pre-defined Gaussian transitions:
\begin{equation}
    p(x_{1:T} | x_0) = \prod_{t=1}^T q(x_t | x_{t-1}),
    \label{eq:diffusion}
\end{equation}
where
\begin{equation}
    q(x_t|x_{t-1}) := \mathcal{N}(x_t; \sqrt{1-\beta_t}x_{t-1}, \beta_t\mathbf{I})
\end{equation}
is the forward Gaussian transition with pre-defined variances $\beta_t$.
Ideally, with large iterative steps $T$ and carefully designed $\beta_t$,
the resulting $x_T$ loses all its information and becomes pure Gaussian noise.
We can marginalize out the intermediate steps and sample at arbitrary timestep $t$
in a closed form:
\begin{equation}
    \begin{aligned}
        q(x_t|x_0) &= \mathcal{N}(x_t; \sqrt{\bar{\alpha_t}}x_0, (1-\bar{\alpha}_t)\mathbf{I}) \\
        \alpha_t &:= 1 - \beta_t, \ \ \bar{\alpha}_t = \prod_{\tau=1}^t \alpha_{\tau}.
    \end{aligned}\label{eq:qxtx0}
\end{equation}
Using the reparameterization trick~\cite{kingma2013auto},
~\cref{eq:qxtx0} can be rewritten as:
\begin{equation}
        x_t = \sqrt{\bar{\alpha}_t}x_0 + \sqrt{1-\bar{\alpha}_t}\epsilon, \ \ \
        x_0 = \frac{x_t - \sqrt{1-\bar{\alpha}_t\epsilon}}{\sqrt{\bar{\alpha}_t}},
        \label{eq:reparam_xt}
\end{equation}
where  $\epsilon\sim\mathcal{N}(0, \mathbf{I})$ is Gaussian noise.
$\sqrt{\bar{\alpha}_t}$ is also referred to as the `noise scale' of $x_t$.
Furthermore, following DDPM ~\cite{ho2020denoising},
we can derive the posterior distribution of $x_{t-1}$ given $x_T$ and $x_0$:
\begin{align}
    q(x_{t-1}|x_t, x_0) &= \mathcal{N}(x_{t-1}; \tilde{\mu}_t(x_t, x_0), \tilde{\beta}_t\mathbf{I}), \label{eq:q_xt1_xt_x0}\\
    \tilde{\mu}_t(x_t, x_0) &:=  \frac{\sqrt{\bar{\alpha}_{t-1}}\beta_t}{1-\bar{\alpha_t}}x_0 + \frac{\sqrt{\alpha}_t(1-\bar{\alpha}_{t-1})}{1-\bar{\alpha}_t}x_t \label{eq:q_xt1_xt_x0_mu} \\
    \tilde{\beta}_t &:= \frac{1-\bar{\alpha}_{t-1}}{1-\bar{\alpha}_t}\beta_t. \label{eq:q_xt1_xt_x0_beta}
\end{align}
The forward posterior in~\cref{eq:q_xt1_xt_x0} will be compared with the learnt
reverse posterior during training.

\subsubsection{Reverse process}
The reverse process (denoising process) learns to recover the original data $x_0$ from Gaussian noise
$x_T\sim\stdgs$ via iterative denoising.
It is formulated as a Markov process with learned transitions:
\begin{equation}
    p_{\theta}(x_{0:T}) = p(x_T)\prod_{t=1}^T p_{\theta}(x_{t-1}|x_t),
\end{equation}
where
\begin{equation}
    p_{\theta}(x_{t-1}|x_t) = \mathcal{N}\big(x_{t-1}; \mu_{\theta}(x_t, t), \Sigma_{\theta}(x_t, t)\big)
    \label{eq:reverse-p}
\end{equation}
is the reverse Gaussian transition with learnt mean $\mu_{\theta}(x_t, t)$ and variance $\Sigma_{\theta}(x_t, t)$.
Note that the variance $\Sigma_{\theta}(x_t, t)$ can be either a time-dependent constant~\cite{ho2020denoising}
or learned by a neural network~\cite{nichol2021improved},
and the mean $\mu_{\theta}(x_t, t)$ is parameterized by a neural network.
The reverse process transforms the standard Gaussian distribution $x^T\sim\stdgs$
into data distribution $p(x_0)$.

With learnt transition distribution $p_{\theta}$, to generate new image from the reverse process,
we first sample $x_T$ from standard Gaussian distribution, and then sample $x_{t-1}$ from $p_{\theta}(x_{t-1}|x_t)$
for $t = T, T-1, ..., 1$.
$x_0$ is the data generated from DDPMs.
Data generation of DDPMs is extremely time-consuming because it
involves hundreds or even thousands of evaluations of
the neural network parameterizing transition $p_{\theta}$.
Therefore, It is necessary to accelerate the sampling process of the DDPM for practical utilization.

\subsubsection{Optimization}
Like other latent variable generative models, such as VAE~\cite{kingma2013auto},
training DDPMs is performed by optimizing the evidence lower bound (ELBO) on
negative log-likelihood~\cite{ho2020denoising}:
\begin{equation}
    \begin{aligned}
        \mathcal{L} &= \mathbb{E}_q \log{\frac{p_{\theta}(x_{0:T})}{q(x_{1:T}|x_0)}} \\
                               &= \mathbb{E}_q \Big[
                                \underbrace{D_{KL}\big(q(x_T|x_0)  \ || \ P(x_T)\big)}_{L_T} \\
                                                &\hspace{1em} + \sum_{t>1} \underbrace{D_{KL}\big(q(x_{t-1}|x_t, x_0) \ || \ p_{\theta}(x_{t-1} | x_t) \big)}_{L_{t-1}} \\
                                                & \hspace{1em} \underbrace{-\log\big(p_{\theta}(x_0|x_1)\big)}_{L_0}
                                                                  \Big] \\
                            &= L_0 + \sum_{t>1} L_t + L_T,
    \end{aligned}\label{eq:ddpm-loss}
\end{equation}
where
\begin{equation*}
    \begin{aligned}
        L_0 &= -\log\big(p_{\theta}(x_0|x_1)\big) \\
        L_t &= D_{KL}\big(q(x_{t-1}|x_t, x_0) \ || \ p_{\theta}(x_{t-1} | x_t) \big) \\
        L_T &= D_{KL}\big(q(x_T|x_0)  \ || \ P(x_T)\big).
    \end{aligned}
\end{equation*}
We refer the readers to Appendix of~\cite{ho2020denoising} for more detailed derivations.
$L_0$ can be evaluated with the histogram of pixel values of images.
$L_T$ is independent of $\theta$ and will ideally be zero with adequate diffusion steps $N$
and proper noise schedule $\{\beta_1, \beta_2, ..., \beta_t\}$.

$L_{t-1}$ in~\cref{eq:ddpm-loss} compares the KL-divergence between estimated posterior
$p_{\theta}(x_{t-1} | x_t)$ and forward posterior in~\cref{eq:q_xt1_xt_x0}.
The KL-divergence between two Gaussians can be analytically expressed:
\begin{equation}
    \begin{aligned}
        L_{t-1} &= D_{KL}\big(q(x_{t-1}|x_t, x_0) \ || \ p_{\theta}(x_{t-1} | x_t) \big) \\
                      &= \frac{1}{2}\log{\frac{\Sigma_{\theta}}{\tilde{\beta}_t}} + \frac{\tilde{\beta}_t^2 + (\tilde{\mu}_t - \mu_{\theta})^2}{2\Sigma_{\theta}^2} - \frac{1}{2}. \\
    \end{aligned}\label{eq:lt1-kld}
\end{equation}

Ho et al. ~\cite{ho2020denoising} suggest to fix the variance $\Sigma_{\theta}$
to a constant value, e.g. $\Sigma_{\theta}=\sigma\mathbf{I}$.
After ignoring all constant variables independent of $\theta$,
~\cref{eq:lt1-kld} can be simplified as:
\begin{equation}
    \begin{aligned}
        L_{t-1} = \mathbb{E}_q \Big[\frac{(\tilde{\mu}_t - \mu_{\theta})^2}{2\sigma^2} \Big].
    \end{aligned}\label{eq:lt1-kld-simp}
\end{equation}
There are different ways to parameterize $\mu_{\theta}$.
The most straightforward option is to approximate $\tilde{\mu}_t$ with a neural network parameterized by $\theta$.
An alternative way is to predict $x_0$ which can be used to derive $\tilde{\mu}_t$ using~\cref{eq:q_xt1_xt_x0_mu}.

The network could also predict the noise $\epsilon$.
Bring ~\cref{eq:reparam_xt} into ~\cref{eq:q_xt1_xt_x0_mu}:
\begin{equation}
    \begin{aligned}
        \tilde{\mu}_t(x_t, x_0) &= \frac{\sqrt{\bar{\alpha}_{t-1}}\beta_t}{1-\bar{\alpha_t}} {\color{blue}{x_0}} + \frac{\sqrt{\alpha}_t(1-\bar{\alpha}_{t-1})}{1-\bar{\alpha}_t}x_t \\
        &= \frac{\sqrt{\bar{\alpha}_{t-1}}\beta_t}{1-\bar{\alpha_t}} \cdot {\color{blue}{\frac{x_t - \sqrt{1-\bar{\alpha}_t\epsilon}}{\sqrt{\bar{\alpha}_t}}}} + \frac{\sqrt{\alpha}_t(1-\bar{\alpha}_{t-1})}{1-\bar{\alpha}_t}x_t \\
        &= \frac{1}{\sqrt{\bar{\alpha}_t}}(x_t - \frac{\beta_t}{\sqrt{1-\bar{\alpha}_t}}\epsilon).
    \end{aligned}\label{eq:mu_xt_x0_reparam}
\end{equation}
We can use a neural network $\epsilon_{\theta}(x_t, t)$ to approximate $\epsilon$ in~\cref{eq:mu_xt_x0_reparam}
and then $\mu_{\theta}$ can be derived by:
\begin{equation}
    \mu_{\theta}(x_t, t) = \frac{1}{\sqrt{\alpha}_t}\big(x_t - \frac{\beta_t}{\sqrt{1-\bar{\alpha}_t}}\epsilon_{\theta}(x_t, t)\big).
\end{equation}
Under this parameterization, the objective $L_{t-1}$ becomes
\begin{equation}
    L_{t-1} = \mathbb{E}_{\epsilon, t}\big[\lVert\epsilon_{\theta}(x_t, x_0, t) - \epsilon\rVert\big]
    \label{eq:loss-discrete}
\end{equation}
where the network predicts the noise $\epsilon$.
Ho et al.~\cite{ho2020denoising} found predicting $\epsilon$ works the best and
the estimated noise $\epsilon_{\theta}$ is equivalent to  Stein score function~\cite{hyvarinen2005estimation,vincent2011connection}
of the log data density.
This connects DDPM with score-based generative models and Langevin dynamics~\cite{song2019generative,song2020improved}.

\subsection{Diffusion Models Conditioned on Noise Level}
In the original DDPM~\cite{ho2020denoising},
noise schedule $\{\beta_1, ..., \beta_T\}$ and number of diffusion (or denoising) steps $N$
have to be carefully tuned to ensure high-quality data generation.
The noise schedule is typically determined by hyper-parameter heuristic, e.g., linear~\cite{ho2020denoising}
or cosine~\cite{nichol2021improved} decay.
To generate high-quality images at high resolution, $N$ has to be large enough as well.
For example, Ho et al. ~\cite{ho2020denoising} use $N=1,000$ to sample $256\times 256$ images.

Instead of conditioning on discrete step index $t$,
Chen et al.~\cite{chen2021wavegrad} reparameterize the model to condition on continuous noise level $\bar{\alpha}_t$.
This allows separate noise schedules $\{\beta_t\}_{t=1}^T$
and number of iterative steps $N$ in training and testing.
Hereby, the network $\epsilon$ is conditioned on noise scale and the objective in~\cref{eq:loss-discrete}
becomes:
\begin{equation}
    L_{t-1} = \mathbb{E}_{\epsilon, t}\big[\lVert\epsilon_{\theta}(x_t, x_0, t) - \epsilon\rVert\big].
    \label{eq:loss-continuous}
\end{equation}

\subsection{Conditional DDPMs for Image SR}

In \imsr{}, we are given a dataset of pairwise low- and \hrim s  $\mathcal{D} = \{(x^{LR}, x^{HR})_1, ..., (x^{LR}, x^{HR})_N\}$
and expected to learn a distribution of high-resolution images conditioned on a low-resolution input image: $p({x^{HR} | x^{LR}})$.


Two concurrent works SR3~\cite{sr3} and SRDiff~\cite{li2022srdiff} approach this problem by adapting the 
denoising diffusion probabilistic models (DDPMs) to conditional image generation.
The basic idea is to use the \lrim{} as the condition for the DDPM image generation framework.
Under this setting, the reverse process of conditional DDPMs is:
\begin{equation}
    p_{\theta}(x_{0:T} | x^{LR}) = p(x_T)\prod_{t=1}^{T} p_{\theta}(x_{t-1}|x_t, x^{LR}).
    \label{eq:reverse-cddpm}
\end{equation}
where the reverse transition $p_{\theta}(x_{t-1}|x_t, x^{LR})$ conditions not only on denoising step $t$, but also
on \lrim{} $x^{LR}$.
Similarly, to generate $x^{HR}$ from $x^{LR}$, we first sample $x_T$ from Gaussian distribution and
then iteratively sample from  $p_{\theta}(x_{t-1}|x_t, x^{LR})$ for $t=T, T-1, ..., 1$.

\section{Partial Diffusion Models for Image SR}\label{sec:partdiff}
In this section, we introduce our proposed DDPM-based method
for \imsr{}.
We first compare the diffusion process of \lrhrim{} in~\cref{sec:diff-lr-hr},
and then introduce the two key components of our proposed method,
partial diffusion and latent alignment,
in ~\cref{sec:part-diff} and~\cref{sec:aug-align}, respectively.

\subsection{Diffusing LR and HR Images}\label{sec:diff-lr-hr}
We first compare the diffusion process of \lrhrim.
Let $p(x_{1:T}^{LR} | x_0^{LR})$ and $p(x_{1:T}^{HR} | x_0^{HR})$ be the forward processes
of \lrhrim,
and $x_t^{LR}$ and $x_t^{HR}$ are the latent states.
The two processes start different distributions, i.e., low- and high-resolution images,
but end up with the same Gaussian distribution $\mathcal{N}(0, \mathbf{1})$.
We hypothesize that the two processes converge at the midway
and $x_t^{HR}$ and $x_t^{LR}$ become indistinguishable.
We verify our hypothesis in two aspects, qualitatively and quantitatively.

First, we visualize the diffusion processes of low- and high-resolution images.
As shown in~\cref{fig:diff-lr-hr}, the diffused images become visually indistinguishable
after several diffusion steps.
Second, we quantitatively analyze the
peak signal noise ratio (PSNR)
between high-resolution image $x_0^{HR}$ and diffused images ($x_t^{LR}$ and $x_t^{HR}$) at different
time steps.
We gather the results of 5,000 low- and high-resolution image pairs
and calculate the average results:
$\text{PSNR}(x_t^{LR}, x_0^{HR})$, $\text{PSNR}(x_t^{HR}, x_0^{HR})$.
Since the diffusion process gradually removes the information from the original image,
the PSNR can reflect how much information is remained during the diffusion process.
\begin{figure}[!h]
    \centering
    \vspace{-2em}
    \begin{overpic}[width=0.7\linewidth]{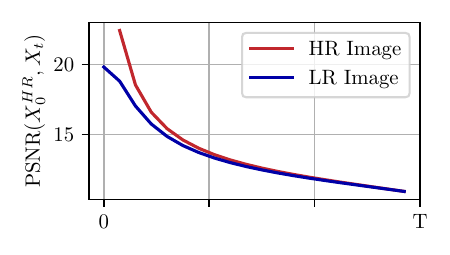}
    \end{overpic}
    \vspace{-2em}
    \caption{
        PSNR between latent states $x_t$ and original high-resolution image  $x_0^{HR}$.
        The diffusion processes of low- and high-resolution images converge at intermediate steps.
    }\label{fig:psnr}
\end{figure}

We can observe from~\cref{fig:psnr} that the latent states of high-resolution images
are more informative than low-resolution counterparts at the beginning, and gradually converge to
the same level after the diffusion process.
The PSNR curves in~\cref{fig:psnr} become very close after about a third of the diffusion steps.
The observation in~\cref{fig:diff-lr-hr} and ~\cref{fig:psnr} reveal that the intermediate latent states
become barely distinguishable in the middle of the diffusion process,
which motivates us to use $p(x_t^{LR}|x_0^{LR})$ as the proxy for $q_{\theta}(x_t^{HR} | x_0^{HR})$
to accelerate data generation.


\subsection{Partial Diffusion Models}\label{sec:part-diff}
Based on the analysis in~\cref{sec:diff-lr-hr}, we propose the partial diffusion Models (PartDiff)
which executes only part of the denoising steps by using
$x_K^{LR}$ as a proxy of $x_K^{HR}$ in the reverse process,
where $K < T$ is an intermediate step after which
$x_K{LR}$ and $x_K{HR}$ become indistinguishable.
PartDiff accelerates the diffusion models by bypassing all steps with $t\ge K$.
Specifically, given the \lrim{} $x_0^{LR}$ as input,
we first diffuse it by $K$ steps to derive $x_K^{LR}$.
This can be analytically evaluated using~\cref{eq:qxtx0}:
$$
q(x_K^{LR}|x_0^{LR}) = \mathcal{N}(x_K; \sqrt{\bar{\alpha_K}}x_0, (1-\bar{\alpha}_K)\mathbf{I}).
$$

Then we use $x_K^{LR}$ as the proxy for $x_K^{HR}$
and start denoising from $x_K^{HR}$ until reach
$x_0^{HR}$, which is the generated \hrim.

\subsection{Latent Alignment}\label{sec:aug-align}
The partial diffusion introduced in~\cref{sec:part-diff} can be seamlessly integrated
into any pretrained diffusion models.
However, the subtle disparity between $x_K^{LR}$ and $x_K^{HR}$
may cause a slight degradation in generation quality due to approximation errors.
The disparity is even more noticeable with larger upsampling factors.
To mitigate the approximation error,
we further propose the `latent alignment'
that gradually aligns the disparity between $x_t^{LR}$ and $x_t^{HR}$.

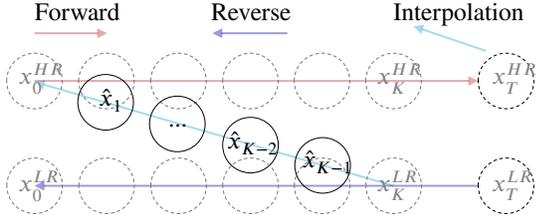
\begin{figure}[!htb]
    \centering
    \vspace{1em}
    \begin{overpic}[width=0.8\linewidth]{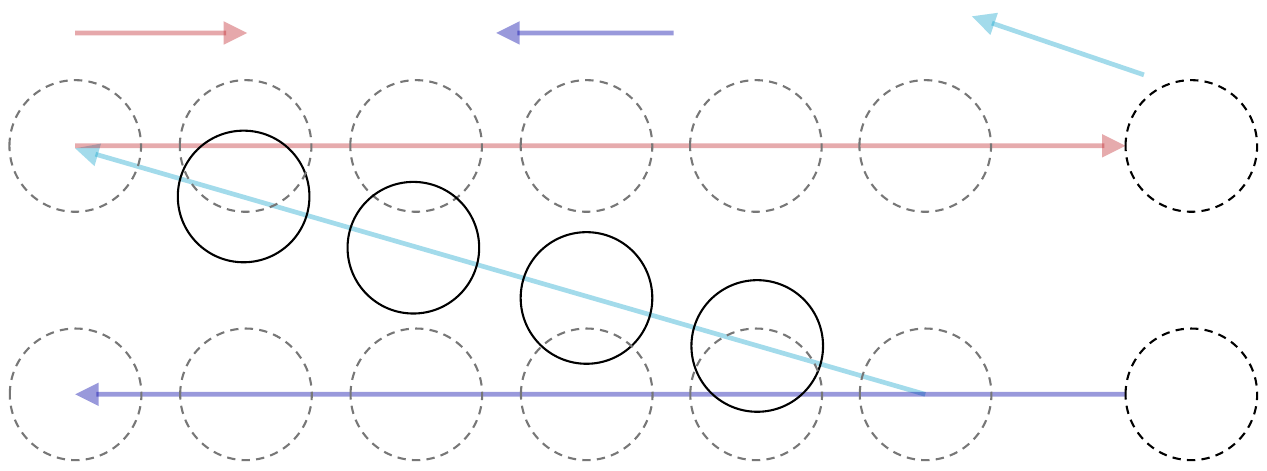}
        \put(6, 37){Forward}
        \put(39, 37){Reverse}
        \put(73, 37){Interpolation}
        \put(91.5,25){{\color{gray}{$x_T^{HR}$}}}
        \put(91.5,5){{\color{gray}{$x_T^{LR}$}}}
        \put(70,25){{\color{gray}{$x_K^{HR}$}}}
        \put(70,5){{\color{gray}{$x_K^{LR}$}}}
        \put(3,25){{\color{gray}{$x_0^{HR}$}}}
        \put(3,5){{\color{gray}{$x_0^{LR}$}}}
        \put(56,9.5){$\hat{x}_{K-1}$}
        \put(42,13){$\hat{x}_{K-2}$}
        \put(31,17){...}
        \put(18,21){$\hat{x}_{1}$}
    \end{overpic}
    \vspace{-1em}
    \caption{
        Schematic illustration of latent interpolation in latent alignment.
        In the reverse (denoising) process, the interpolated latent gradually
        shifts from low-resolution to high-resolution latent states.
    }
    \label{fig:aug-align}
    \vspace{-1em}
\end{figure}

For each training iteration, we first randomly sample a step-index $t\in(0, K]$
and diffuse both \lrhrim{} to derive
$q(x_t^{LR}|x_0^{LR})$ and $q(x_t^{HR}|x_0^{HR})$ according to~\cref{eq:qxtx0}.
Then, we linearly interpolate between them:
\begin{equation*}
    q(\hat{x}_t | x_0^{LR}, x_0^{HR}) = 
    \mathcal{N}\Big(
        \hat{x}_t; \sqrt{\bar{\alpha_t}}\big(\lambda x_0^{HR} + (1-\lambda)x_0^{LR}\big), (1-\bar{\alpha}_t)\mathbf{I}
    \Big),
\end{equation*}
where $\hat{x}_t$ is the interpolated latent and
$\lambda=\frac{K-t}{K} \in [0, 1]$
is a weight linearly increasing from 0 to 1.
~\cref{fig:aug-align} illustrates the latent interpolation in latent alignment.
Intuitively, latent alignment gradually distributes the disparity
to each step and avoids a sudden large gap at $t=K$.

Similar to~\cref{eq:q_xt1_xt_x0}, the forward diffusion posterior for $t<K$ is:
\begin{equation}
    \begin{aligned}
    q(\hat{x}_{t-1} | x_t, x_0^{LR}, x_0^{HR}) &= \mathcal{N}\big(x_{t-1}; \hat{\mu}_t(x_t, x_0), \hat{\beta}_t\mathbf{I}\big), \\
\end{aligned}\label{eq:qxt1_xtx0lrx0hr}
\end{equation}
where
\begin{equation*}
\begin{aligned}
    \hat{\mu}_t(x_t, x_0) &= \lambda\tilde{\mu}_t(x^{HR}_t, x^{HR}_0) + (1-\lambda)\tilde{\mu}_t(x^{LR}_t, x^{LR}_0) \\
    \hat{\beta}_t &= \tilde{\beta}_t.
\end{aligned}
\end{equation*}
and $\tilde{\mu}(\cdot, \cdot)$ and $\hat{\beta}_t$ are the mean and variance of forward posterior
defined in~\cref{eq:q_xt1_xt_x0_mu} and ~\cref{eq:q_xt1_xt_x0_beta}, respectively.
The interpolated posterior is used as the target distribution in training the denoising model,
and the loss term $L_{t-1}$ in~\cref{eq:ddpm-loss} becomes
$$
L_{t-1} = D_{KL}\Big(q(\hat{x}_{t-1} | x_t, x_0^{LR}, x_0^{HR}) \ || \ p_{\theta}(x_{t-1} | x_t)\Big)
$$
and the denoising model learns to gradually approach $X_0^{HR}$ from $X_K^{LR}$,
as illustrated in~\cref{fig:aug-align}.

\section{Experiments}\label{sec:exp}
In this section, we introduce our detailed implementation
and report the experimental results.
In~\cref{sec:impl-details}, we introduce the implementation details
in our experiments, including datasets, model architecture, and model training.
In~\cref{sec:exp-mri,sec:exp-natural-img}, we report experimental results on
MRI and natural images
and make qualitative and quantitative comparisons with other \sr{} methods.
In addition to \sr{}, in~\cref{sec:exp-downstream}, we apply
the \sr{} methods to several downstream tasks.
Finally, we compare our method with SR3~\cite{sr3} in~\cref{sec:comp-sr3}.

\subsection{Implementation details}\label{sec:impl-details}
All the models have been implemented with the PyTorch~\cite{paszke2019pytorch} framework.
We use $T=2,000$ in training all the models, and the inference denoising steps are
detailed in~\cref{sec:exp-noise-schedule}.
All hyper-parameters, unless otherwise stated, are kept identical with that of SR3~\cite{sr3} for fair comparisons.

\subsubsection{Network Architecture and Model Training}
We use a similar UNet architecture to SR3~\cite{sr3} which is adopted from DDPM~\cite{ho2020denoising}
with several minor modifications.
We change the number of residual blocks in each down-sample stage
from 3 to 2 for $64\times64\rightarrow 512\times512$ \sr{} to reduce the model size.
Instead of using a fixed number of groups, which is 32 in SR3,
we fix the number of channels in each group to 16.
Detailed network configurations for each task is summarized in~\cref{tab:netarch},
where base channels are the feature dimension in the first downsampling stage,
channels multiplier is the multiplicative factor of the feature dimension of subsequential stages,
and residual blocks are the number of residual blocks in each stage.

\begin{table}[!htb]
    \centering
    \caption{
        Detailed network configurations for each task.
        The number inside $(\cdot)$ denotes the target resolution.}
    \vspace{-1em}
    \resizebox{1\linewidth}{!}{
    \begin{tabular}{cccc}
        \Xhline{2\arrayrulewidth}
        Modality & \makecell{Base \\ Channels} & \makecell{Channels \\ Multiplier} & \makecell{Residual \\ Blocks} \\
        \hline
        MRI (*)   & 128 & \{1, 1, 2, 2, 4, 4\} & 3 \\
        RGB (128, 256)   & 128 & \{1, 1, 2, 2, 4, 4\} & 3 \\
        RGB (512)   & 64 & \{1, 1, 2, 2, 4, 4, 8, 8\} & 2 \\
        \hline
    \end{tabular}
    }
    \label{tab:netarch}
\end{table}

We upsample the \lrim{} to the target resolution using bicubic interpolation with anti-aliasing enabled,
and the upsampled image is concatenated with Gaussian noise $x^{HR}_T$
in the channel dimension as the input to the model.

All models are trained on a server with 4 NVIDIA RTX 3090 GPUs.
We use the AdamW optimizer~\cite{loshchilov2018decoupled} with a fixed learning rate of $10^{-4}$ and zero weight decay.
RGB and MRI image models are trained for 0.5M and 0.2M iterations, respectively.
The batch size on each GPU is set to 4 for all experiments except
$64\times64\rightarrow512\times512$ facial \imsr{}  where the batch size is 2 due to limited GPU memory.
The training recipe is adopted from previous studies~\cite{sr3,ho2020denoising}
with minor modifications to model size, batch size, and training iterations because
of our computational budget.
Detailed configurations are summarized in~\cref{tab:batch-iters}.
\begin{table}[!htb]
    \centering
    \vspace{-1.5em}
    \caption{
        Batch size (BS) and total training iterations for each setting.
    }
    \vspace{-1em}
    \resizebox{0.75\linewidth}{!}{
    \begin{tabular}{llcc}
        \Xhline{2\arrayrulewidth}
        \makecell{Target \\ resolution} & Dataset & \makecell{Batch \\ size} & Iters (M) \\
        \hline
        $128\times 128$   &  \multirow{3}{*}{FFHQ~\cite{karras2019style}} & 32 & 0.5 \\
        $256\times 256$ &   & 32 & 0.5 \\
        $512\times 512$  &   & 8 & 0.75 \\
        $256\times 256$  &  ImageNet~\cite{deng2009imagenet} & 32 & 1 \\
        $320\times 320$  &  Prostate MRI & 64 & 0.25 \\
        \hline
    \end{tabular}
    }
    \label{tab:batch-iters}
\end{table}

\subsubsection{Noise Schedule and Denoising Steps}\label{sec:exp-noise-schedule}
\begin{figure}[!htb]
    \centering
    \vspace{-0.5em}
    \begin{overpic}[width=0.85\linewidth]{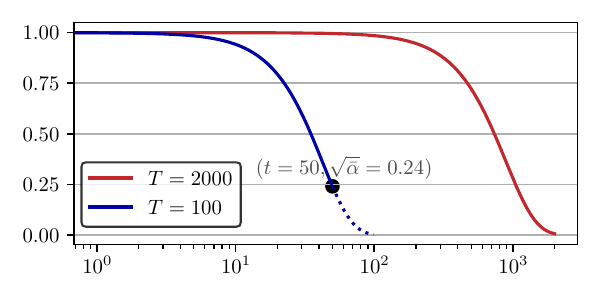}
    \end{overpic}
    \vspace{-2em}
    \caption{
        The plot of noise scales $\sqrt{\bar{\alpha}}$ with different noise schedules.
        The noise scale for $T=100$ and $T=50$ are manually tuned.
    }\label{fig:noise-scale}
\end{figure}

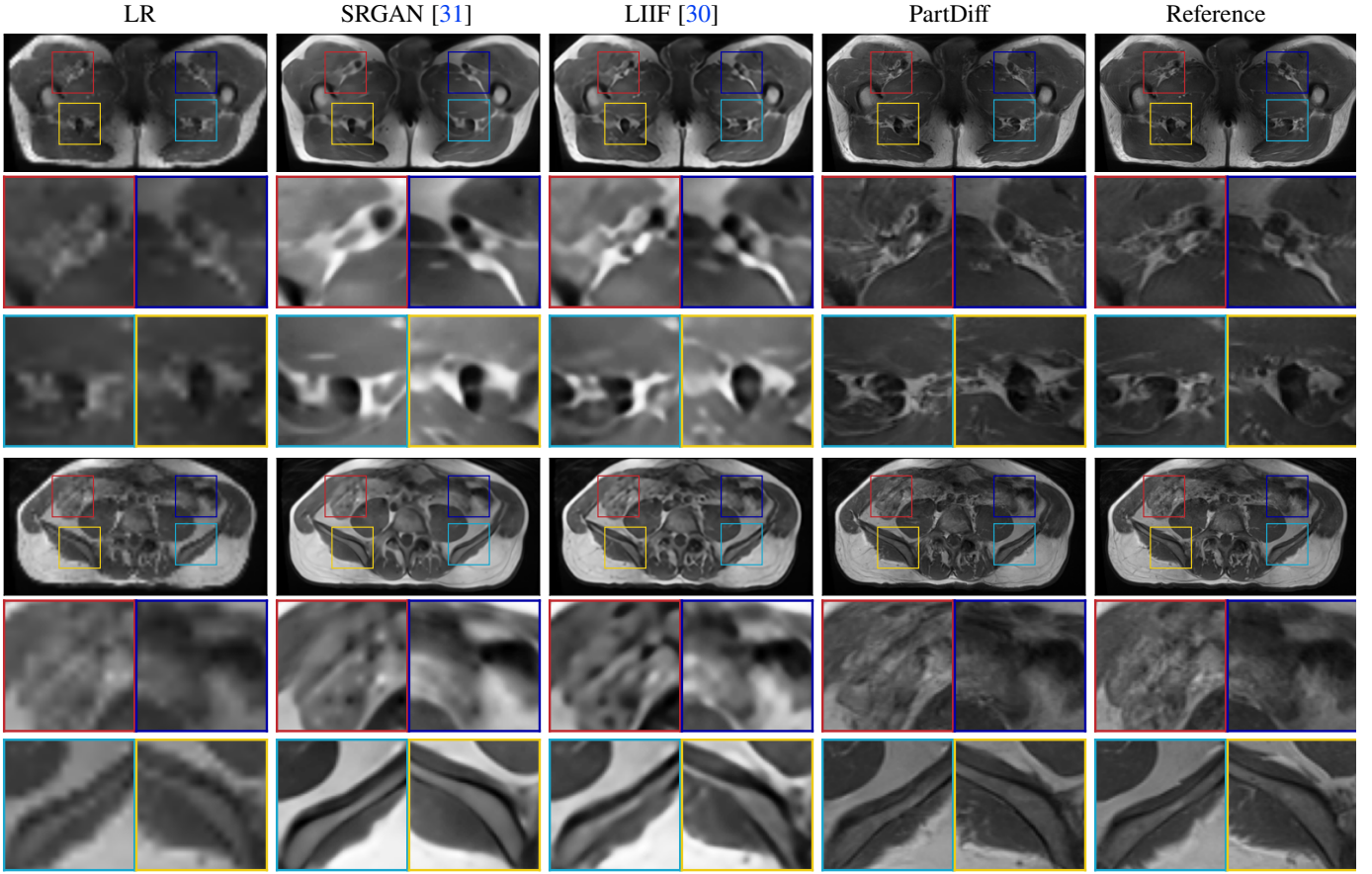
\begin{figure*}[!htb]
    \vspace{1em}
    \begin{overpic}[width=1\linewidth,right]{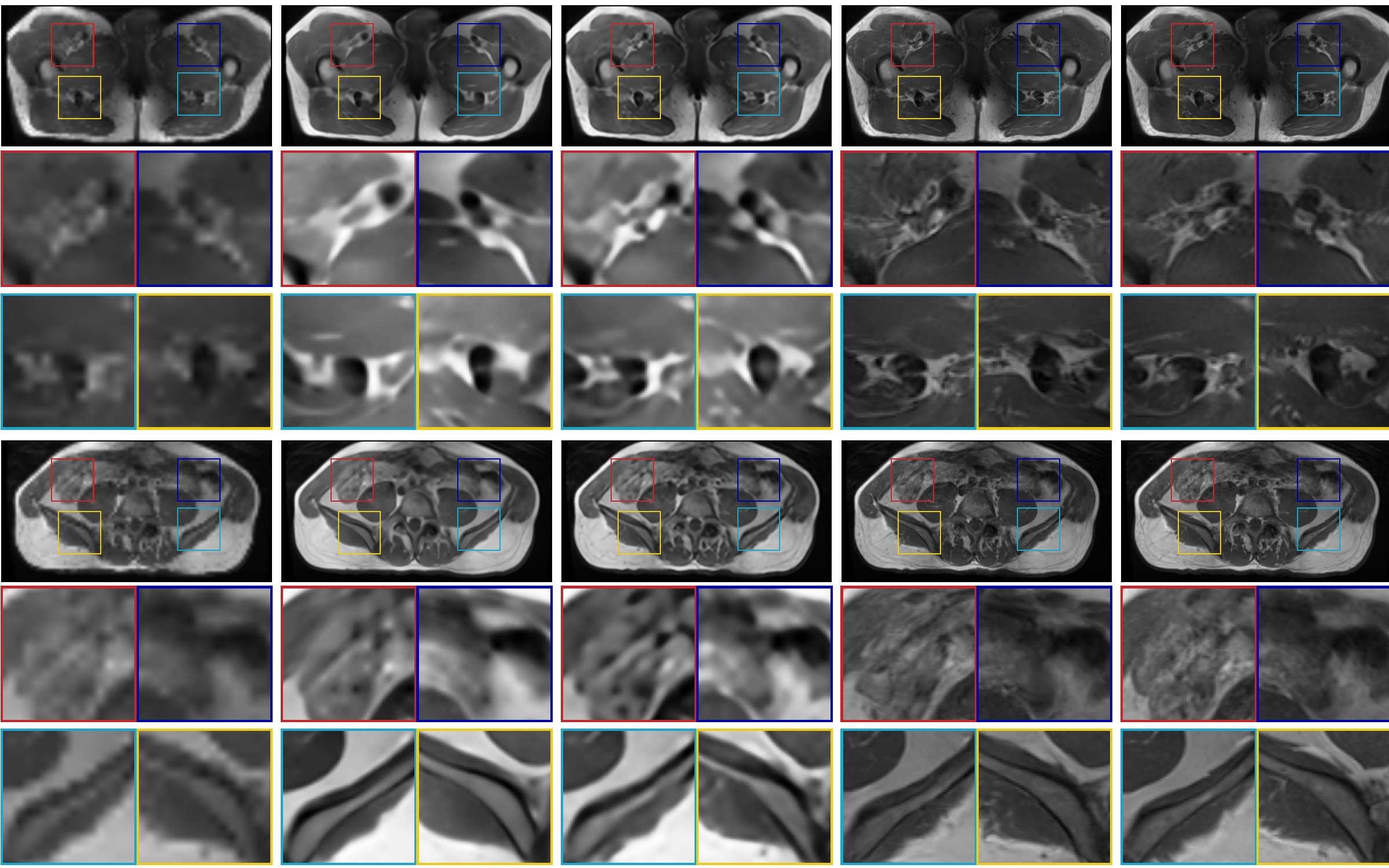}
        \put(9,63){LR}
        \put(25,63){SRGAN~\cite{ledig2017photo}}
        \put(46,63){LIIF~\cite{chen2021learning}}
        \put(67,63){PartDiff}
        \put(86,63){Reference}
    \end{overpic}
    \vspace{-2.5em}
    \caption{
        Super-resolution results of ($80\times80\rightarrow320\times320$) T1-weighted MRI images.
        Some of the areas are enlarged to visualize the fine details.
        Our method generates high-frequency textures, and the overall contrast
        is the closest to the high-resolution images compared to other approaches.
    }
    \vspace{-1em}
    \label{fig:t1w-idx}
\end{figure*}
Follow DDPM~\cite{ho2020denoising} and SR3~\cite{sr3},
we set the forward process variances $\{\beta_t\in(0, 1)\}_{t=1}^{T}$ to constants increasing linearly from
$\beta_1$ to $\beta_T$, defined as Linear$(\beta_1, \beta_T, T)$.
We set $T=2,000$ and $\beta_1=5\times 10^{-5}, \beta_T=0.01$ for all experiments.

During training, we adapt the technique from~\cite{chen2021wavegrad}
and condition the model on the continuous noise scale
$\sqrt{\bar{\alpha}_t}$ rather than discrete iteration indice $t$.
Specifically, at each training step, we randomly sample a discrete iteration index $t$ from uniform
distribution $\mathbf{U}(1, T)$, and then
randomly sample the noise scale $\sqrt{\bar{\alpha}}$ uniformly from $\mathbf{U}(\sqrt{\bar{\alpha}_t}, \sqrt{\bar{\alpha}_{t-1}})$.
By conditioning on continuous noise scales, the mode is more flexible in
a number of diffusion steps and the noise schedule during inference.

During inference, we set $T=100$
and manually search a linear noise schedule.
Concretely, with a pretrained model, we do a grid search on $\beta_1$ and $\beta_T$ on the test set
to find a proper noise scale.
Consequently, the inference noise schedule used in our experiment is Linear$(10^{-5}, 0.213, 100)$.
The noise schedules in training and inference are plotted in~\cref{fig:noise-scale}.

Given a specific inference noise schedule Linear$(\beta_0, \beta_T, T)$,
PartDiff can easily bypass some of the denoising steps by setting up a noise scale
threshold.
For example, with inference noise schedule Linear$(10^{-5}, 0.213, 100)$
({\color{blue}{blue}} curve in~\cref{fig:noise-scale}),
we set the threshold to $\sqrt{\bar{\alpha}_{K=50}}=0.25$ and this reduces the denoising
steps from 100 to 50.
Note that $K$ is closely related to image modality and upsampling factors.
In general, high-upsampling factors and images with high-frequency details
require a larger $K$ (and more denoising steps).



\subsubsection{Data and Data Preprocessing}
We test our method on clinical MRI images and natural images.
We crop the images into squares and then resize them into proper sizes before feeding them into the model.
The training images are randomly cropped along the longer edge and test images are center cropped.
A random left-right flip is performed for data augmentation.
SR3~\cite{sr3} suggests applying varying amounts of Gaussian filters to the low-resolution images
during training and proves it improves the quality of generation significantly.
We adopt this trick and the sigma of the Gaussian kernel is randomly chosen from
$(0\sim 1.5)$.

\mypar{Prostate MRI}
For MRI, we experiment on two datasets:
1) our in-house prostate MRI dataset, and
2) the publically available prostateX dataset from the cancer imaging archive (TCIA)~\cite{clark2013cancer}.
Our in-house dataset consists of T1- and T2-weighted images from 871 patients.
We randomly select 200 patients for testing and use other patients for training.
We collect all slices of different scan planes (axial, sagittal, and coronal)
and totally we have 18,813 T1-weighted slices and  17,445 T2-weighted slices.
The prostateX dataset consists of 347 patients in total. 
203 patients are randomly selected for the test, and the rest of the patients are used as a training set.

\begin{figure*}[!htb]
    \vspace{1em}
    \begin{overpic}[width=0.98\linewidth,right]{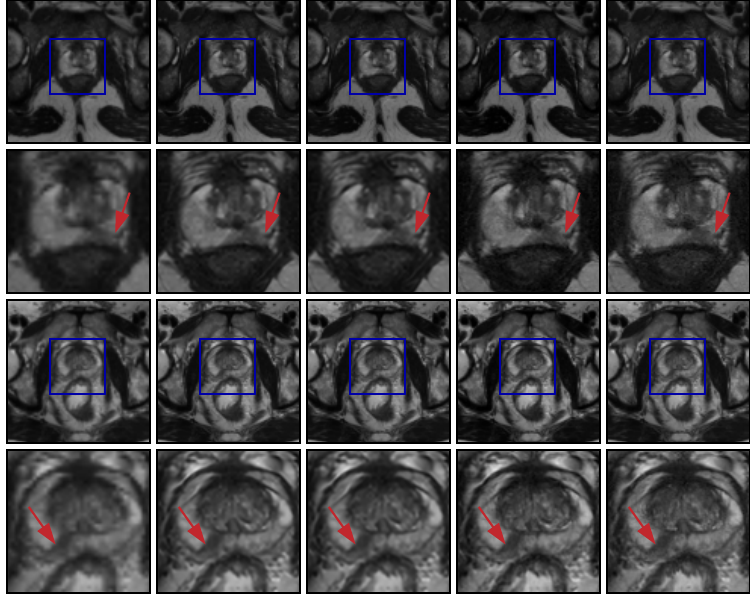}
        \put(10,79){LR}
        \put(27,79){SRGAN~\cite{ledig2017photo}}
        \put(47,79){LIIF~\cite{ledig2017photo}}
        \put(67,79){PartDiff}
        \put(87,79){HR}
    \end{overpic}
    \vspace{-2.5em}
    \caption{
        Some $80\times80\rightarrow320\times320$ \sr{} results on T2-weighted prostate MRI images.
        The prostate area is zoomed-in to show finer details and lesions.
        Lesions are marked with red arrows in the enlarged patches.
    }
    \vspace{-1em}
    \label{fig:t2w-idx}
\end{figure*}
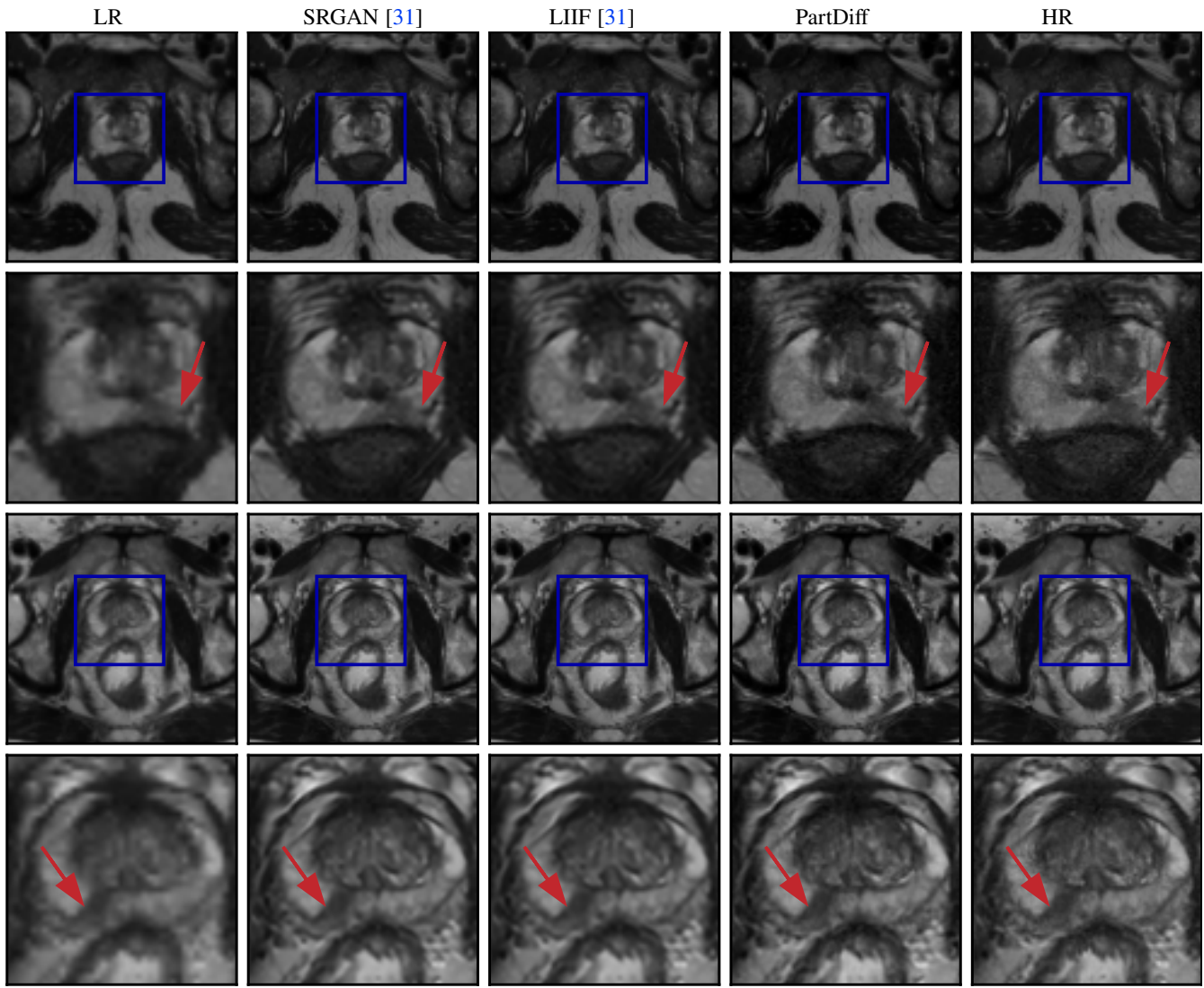

\mypar{Natural images}
In addition to MRI, we test our method on two kinds of natural images:
facial images and the ImageNet~\cite{deng2009imagenet}.
Following SR3~\cite{sr3},
we train facial \imsr{} models on Flickr-faces-HQ (FFHQ)~\cite{karras2019style}
dataset and evaluate on the CelebA-HQ~\cite{karras2018progressive} dataset.
The upsampling scale is set to $8\times$ to test the performance under radical situations.
We test two different resolutions: $16\times16\rightarrow128\times128$ and $64\times64\rightarrow512\times512$.
The ImageNet dataset has more than one million images of 1,000 categories.
We use its training set model training and test on its validation set.
We perform $\times4$ ($64\times64\rightarrow256\times256$) \sr{} on this dataset.

\subsection{Experiments on Clinical MRI}\label{sec:exp-mri}
We first test our method on clinical prostate MRI scans.
%
Let $(h\times w\times d)$ be the shape of a multiple slices of 2D MRI images where  $(h\times w)$ is the in-plane
resolution and $d$(epth) is the number of slices.
We perform \sr{} on two different settings:
1) in-plane \sr{} that improves the resolution in the $(h\times w)$ plane
and 2) through-plane \sr{} that improves the resolution in the $d$-axis.

\subsubsection{In-plane MRI Super-resolution}
We use each 2D MRI image as a training sample and downsample
the images to get high- and low-resolution pairs for in-plane  MRI \sr.
We conduct MRI \sr{} on both T1- and T2-weighted images
as they are often used to identify prostate cancers
following the clinical guideline ~\cite{Turkbey2019PIRADS}.
An example of the prostate cancer lesion is shown in ~\cref{fig:t2w-idx}
as asymmetrical hypointensity areas in the T2-weighted images.

\mypar{Qualitative comparisons}
Some visual results are shown in~\cref{fig:t1w-idx,fig:t2w-idx}.
~\cref{fig:t1w-idx} shows the \sr{} results on $80\times80\rightarrow320\times320$ \sr
of T1-weighted MRI images with zoomed-in patches.
SRGAN~\cite{ledig2017photo} generates decent outputs, but the contrast is obviously
different from the high-resolution images (see zoomed-in patches).
Both LIIF~\cite{chen2021learning} and SRGAN~\cite{ledig2017photo} generate blurry results without high-frequency textures.
Our method generates realistic textures and the overall contrast is the closest
to high-resolution references compared to other approaches.

~\cref{fig:t2w-idx} presents some \sr{} results on $160\times160\rightarrow320\times320$ \sr{} of T2-weighted MRI images.
The prostate area is marked by a blue square and enlarged to visualize the detailed prostate cancer lesion inside the prostate.
Prostate tumors are marked with red arrows.
Both the GAN-based method (SRGAN) and implicit representation-based method (LIIF)
deteriorate the contrast between the lesion and normal tissue, making the lesion less
clear in the image.
Our method preserves the contrast the most, and the lesion is more evident in the image than the other methods.

\mypar{Quantitative performance}
~\cref{tab:prostatex} and ~\cref{tab:idx} summarize the quantitative performance on ProstateX and
our in-house prostate MRI datasets, respectively.
We use the peak signal-to-noise ratio (PSNR) and structural similarity index measure (SSIM)
to measure the similarity between upsampled images and high-resolution references.
As shown in ~\cref{tab:prostatex} diffusion-based methods outperform other methods
and the performance gap is even more clear with larger upsampling factors.
Partial diffusion models achieve very close performance to SR3 with significantly fewer denoising steps.%
From ~\cref{tab:idx} we can see that the PSNR and SSIM are generally higher in T1-weighted images
than T2-weighted images.
This is probably because T1-weighted images are less noisy than T2-weighted images,
as shown in~\cref{fig:t1w-idx,fig:t2w-idx}.
Diffusion-based methods achieve state-of-the-art performance on both T1-weighted
and T2-weighted images and our method performs on par with SR3 with less
computational footprint.


\begin{figure*}[!htb]
    \vspace{2em}
    \begin{overpic}[width=1\linewidth]{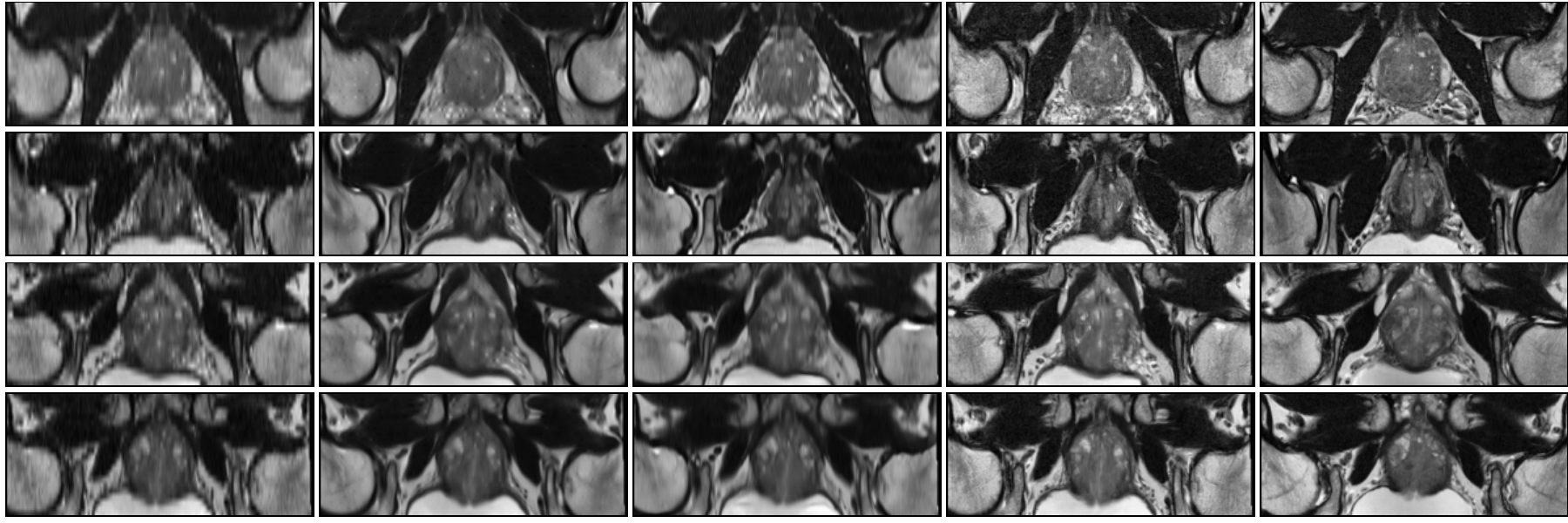}
        \put(9, 35.4){Input }
        \put(5,33.7){($d\times w$), axial}
        \put(24, 34.5){SRGAN~\cite{ledig2017photo,sood2019anisotropic} }
        \put(46, 34.5){LIIF~\cite{chen2021learning}}
        \put(66, 34.5){PartDiff}
        \put(83, 35.2){Visual reference}
        \put(83, 33.7){($h\times w$), coronal}
    \end{overpic}
    \vspace{-2.5em}
    \caption{
        Example results on through-plane MRI image \sr.
        The model is trained with in-plane slices $(h\times w)$ of coronal scan
        and the test input is the through-plane $(w\times d)$ images of  axial scan.
        The visual reference is an in-plane slice ($h\times w$) from
        coronal scan and is not necessarily aligned with the results.
    }
    \vspace{-1.5em}
    \label{fig:through-plane}
\end{figure*}

\begin{figure}[!tb]
    \begin{overpic}[width=1\linewidth]{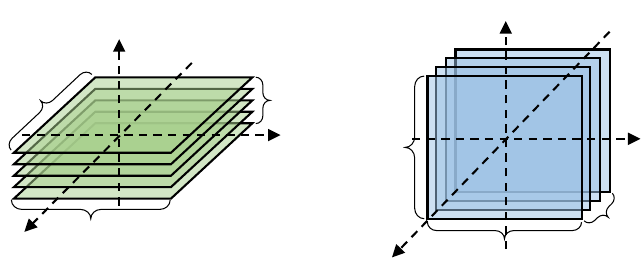}
        \put(12, 37.5){{\color{gray}{Axial scan}}}
        \put(70, 38.5){{\color{gray}{Coronal scan}}}
        %
        \put(1,1){AP}
        \put(20,32){SI}
        \put(40,15){LR}
        \put(42,23){$d$}
        \put(14,4){$w$}
        \put(4,24){$h$}
        \put(55,1){AP}
        \put(80,35){SI}
        \put(97,15){LR}
        \put(60,16){$h$}
        \put(74,0){$w$}
        \put(95,5){$d$}
    \end{overpic}
    \vspace{-2em}
    \caption{
        Axial scan, coronal scan, and the
        three anatomical directions:
        anterior$\leftrightarrow$posterior (AP),
        left$\leftrightarrow$right (LR) and
        superior$\leftrightarrow$inferior (SI).
    }\label{fig:cor-ax}s
\end{figure}
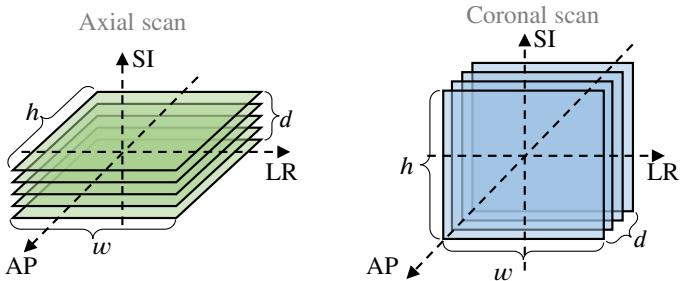

\begin{table}[!htb]
    \centering
    \vspace{-1em}
    \caption{
        In-plane MRI \sr{} results on ProstateX dataset.
    }
    \vspace{-1em}
    \newcommand{\CC}{\cellcolor{gray!15}}
    \setlength{\tabcolsep}{1.6pt}
    \resizebox{0.8\linewidth}{!}{
        \begin{tabular}{lcc|cc}
            \Xhline{2\arrayrulewidth}
             & \multicolumn{2}{c|}{$\times2$} & \multicolumn{2}{c}{$\times4$} \\
            Method & PSNR ($\uparrow$) & SSIM ($\downarrow$) & PSNR ($\uparrow$) & SSIM ($\downarrow$) \\
            \hline
            Bicubic  & 32.0775 & 0.9226 & 24.6771 & 0.6808 \\
            Regression & 32.7084 & 0.9109 & 25.3750 & 0.7112 \\
            SRGAN~\cite{ledig2017photo} & 33.3979 & 0.9135 & 26.9268 & 0.7503 \\
            LIIF~\cite{chen2021learning} & 33.4791 & 0.9270 & 27.9112 & 0.7929 \\
            \CC SR3 ($T=100$) & \CC\textbf{33.5382} & \CC0.9301 & \CC 28.2462 & \CC\textbf{0.8101} \\
            \CC PartDiff ($K=25$) & \CC 33.5194 & \CC 0.9287 & \CC 28.2329 & \CC 0.8002\\
            \CC PartDiff ($K=50$) & \CC 33.5318 & \CC 0.9295 & \CC \textbf{28.2463} & \CC 0.8093\\
            \Xhline{2\arrayrulewidth}
        \end{tabular}
    }\label{tab:prostatex}
\end{table}

\begin{table}[!htb]
    \centering
    \vspace{-1em}
    \caption{
        Quantitative performance of in-plane MRI
        \sr{} results on our in-house prostate MRI dataset.
    }
    \vspace{-1.2em}
    \newcommand{\CC}{\cellcolor{gray!15}}
    \setlength{\tabcolsep}{1.2pt}
    \resizebox{0.85\linewidth}{!}{
    \begin{tabular}{lc|cc|cc}
        \Xhline{2\arrayrulewidth}
        \multirow{2}{*}{Method} & & \multicolumn{2}{c|}{T1W} & \multicolumn{2}{c}{T2W}\\
        & & PSNR ($\uparrow$) & SSIM ($\uparrow$) & PSNR($\uparrow$) & SSIM ($\uparrow$) \\
        \hline
        Bicubic & \multirow{6}{*}{$\times2$} & 33.3657 & 0.7509 & 29.5263 & 0.9091 \\
        Regression & & 34.4634 & 0.9265 & 32.7347 & 0.9035 \\
        SRGAN~\cite{ledig2017photo} & & 36.0610 & 0.9434 & 36.2626 & 0.9365 \\
        LIIF~\cite{chen2021learning}  & & 37.3901 & 0.9524 & 35.8739 & 0.9433\\
        SR3~\cite{sr3} ($T=100$) & & \CC \textbf{37.5087} &  \CC 0.9670 & \CC \textbf{36.3017} & \CC \textbf{0.9413} \\
        PartDiff ($K=25$) & & \CC 37.5109 &  \CC 0.9672 & \CC 36.2859 & \CC 0.9411 \\
        PartDiff ($K=50$) & &  \CC  37.5071 &  \CC \textbf{0.9672} & \CC 36.2770 & \CC 0.9411 \\
        \hline
        Bicubic & \multirow{6}{*}{$\times4$} & 26.5617 & 0.8278 & 24.9121 & 0.6999 \\
        Regression & & 29.0838 & 0.8095 & 27.3084 & 0.7199\\
        SRGAN~\cite{ledig2017photo} & & 30.2013 &  0.8444 & 29.6069 & 0.7443 \\
        LIIF~\cite{chen2021learning}  & & 31.3574 & 0.8647 & 30.2751 & 0.8261 \\
        SR3~\cite{sr3} ($T=100$) & & \CC 31.5104 &  \CC \textbf{0.8656} & \CC \textbf{30.4349} & \CC 0.8348 \\
        PartDiff ($K=25$) & & \CC 31.3581 & \CC 0.8607 & \CC 30.3109 & \CC 0.8301 \\
        PartDiff ($K=50$) & & \CC \textbf{31.5149} &  \CC 0.8617 & \CC  30.1342 & \CC \textbf{0.8156} \\
        \Xhline{2\arrayrulewidth}
    \end{tabular}
    }
    \label{tab:idx}
    \vspace{-1.2em}
\end{table}

\subsubsection{Through-plane MRI Super-resolution}\label{sec:exp-through}
Multi-slice 2D MRI images have an anisotropic resolution due to the
lower through-plane resolution.
For example, in our prostate MRI dataset, the in-plane pixel spacing
(the physical distance between two pixels) is 0.625 mm, while the slice spacing
(the physical distance between two slices) is 3.6 mm.
In this experiment, we test our method in improving the through-plane resolution of MRI,
where the upsampling scale is set to $6=0.36/0.625$.

To construct low- and high-resolution training pairs,
we use two distinct scans acquired in orthogonal planes, e.g., axial scan and coronal scan,
during training and testing.
The two orthogonal scans are illustrated in~\cref{fig:cor-ax}.
Let anterior$\leftrightarrow$posterior (AP) 
left$\leftrightarrow$right (LR) and 
superior$\leftrightarrow$inferior (SI) 
be the anatomical directions in 3D space.
Axial scan captures 2D slices in the AP and LR plane, and coronal scan captures slices in
SI and LR plane.
Let $h\times w\times d$ be the shape of MRI data where $h\times w$ is the in-plane image size
and $d$ is the number of slices.
In our data, $h=w=320$ and $d=20$.
During training, we collect $h\times w$ in-plane slices from the 
coronal scan 
as the high-resolution image,
and downsample those slices along the $h$ dimension (SI) to simulate
the low through-plane (SI direction) in the axial scan.
The models learn to \sr{} along the SI direction
by training with the data pairs.
During test, the model takes as input the $w\times d$ through-plane slices of the
axial scan and performs \sr{} in the $d$ dimension (SI).

\begin{figure*}[!htb]
    \vspace{1.5em}
    \begin{overpic}[width=1\linewidth]{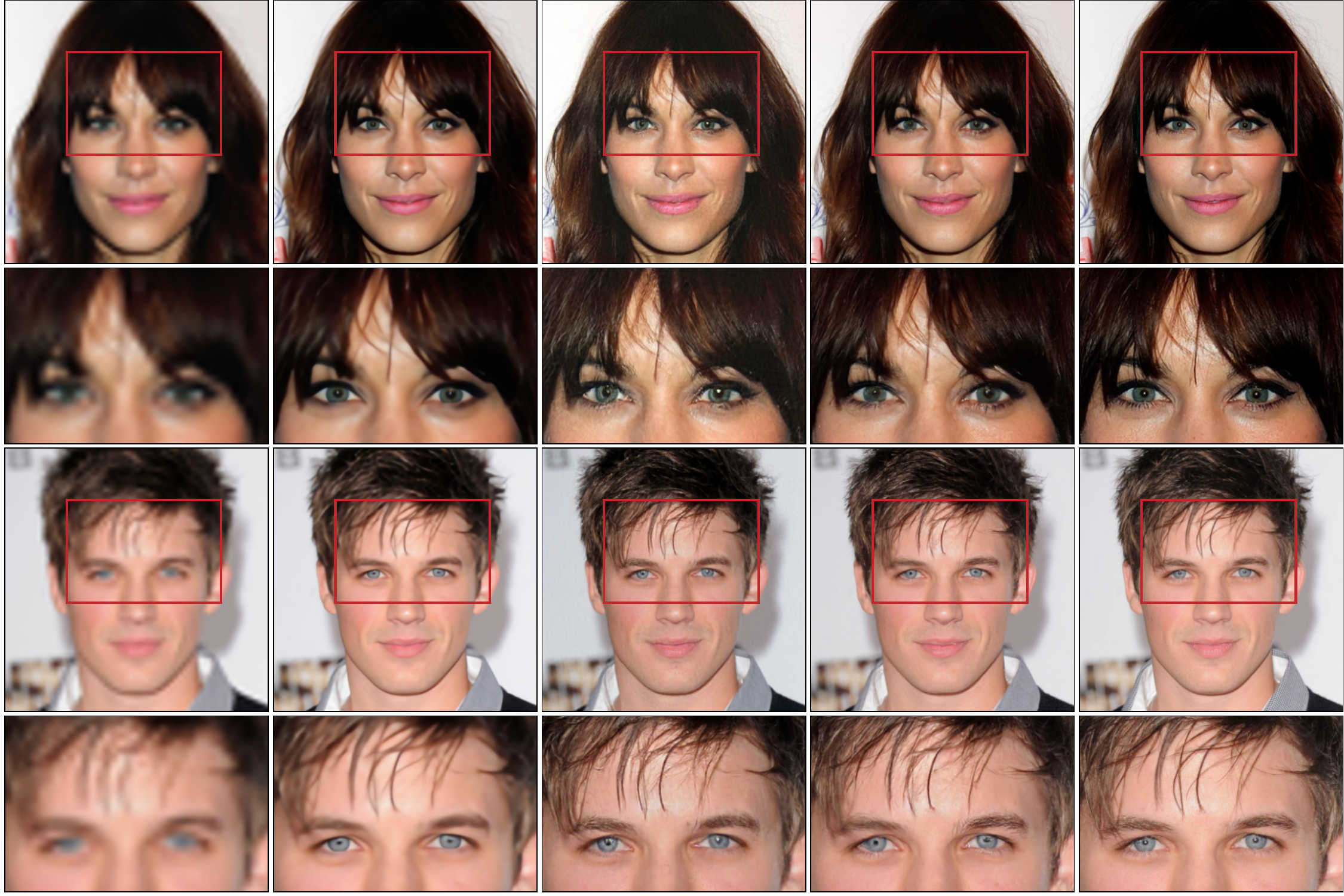}
        \put(7,67.5){Bicubic}
        \put(26,67.5){Regression}
        \put(46,67.5){SR3~\cite{sr3}}
        \put(67,67.5){PartDiff}
        \put(86,67.5){Reference}
    \end{overpic}
    \vspace{-2.5em}
    \caption{
        Example results ($64\times64\rightarrow 512\times512$) on the CelebA dataset
        with enlarged patches showing the details.
        \ifdefined\withappendix
        More results can be found in the Appendix.
        \fi
    }
    \label{fig:face512}
\end{figure*}

\begin{figure}[!htb]
    \vspace{1.5em}
    \begin{overpic}[width=1\linewidth]{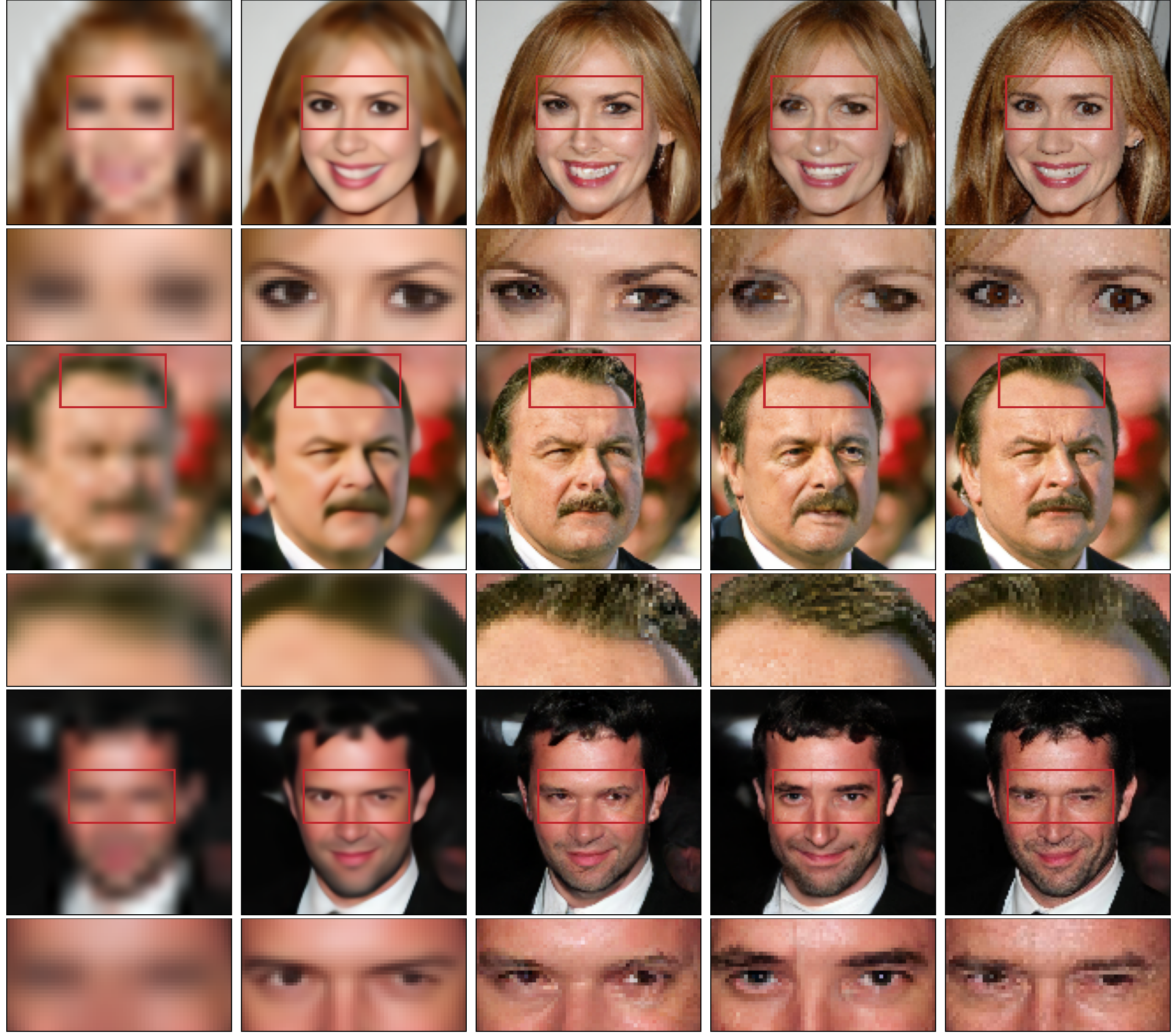}
        \put(5,88.8){Bicubic}
        \put(22,88.8){Regression}
        \put(43,88.8){SR3~\cite{sr3}}
        \put(63,88.8){PartDiff}
        \put(82,88.8){Reference}
    \end{overpic}
    \vspace{-2.5em}
    \caption{
        Example results ($16\times16\rightarrow 128\times128$) on the CelebA dataset.
        \ifdefined\withappendix
        More results can be found in the Appendix.
        \fi
    }\label{fig:face128}
    \vspace{-2.5em}
\end{figure}

Example through-plane MRI \sr{} results of the axial scan are shown in~\cref{fig:through-plane}.
There is no such `ground-truth' in through-plane \sr,
we use the in-plane slice from a separate acquisition of
the coronal scan as the visual reference.
The visual references are not necessarily perfectly aligned with the \sr{} results due to
potential patient motion between the two scans.
Note that we use an average of 6 ($\approx 3.6/0.625$) consecutive through-plane
images as the input to \sr{} models to compensate for the differences
in pixel spacing (0.625 mm) and the slice thickness (3.6 mm).
As shown in~\cref{fig:through-plane}, SRGAN generates blurry results which do not contain much
texture details.
LIIF and our method generate much better high-frequency details and are generally aligned better
with the visual reference.

\begin{figure*}[!htb]
    \vspace{1em}
    \begin{overpic}[width=1\linewidth]{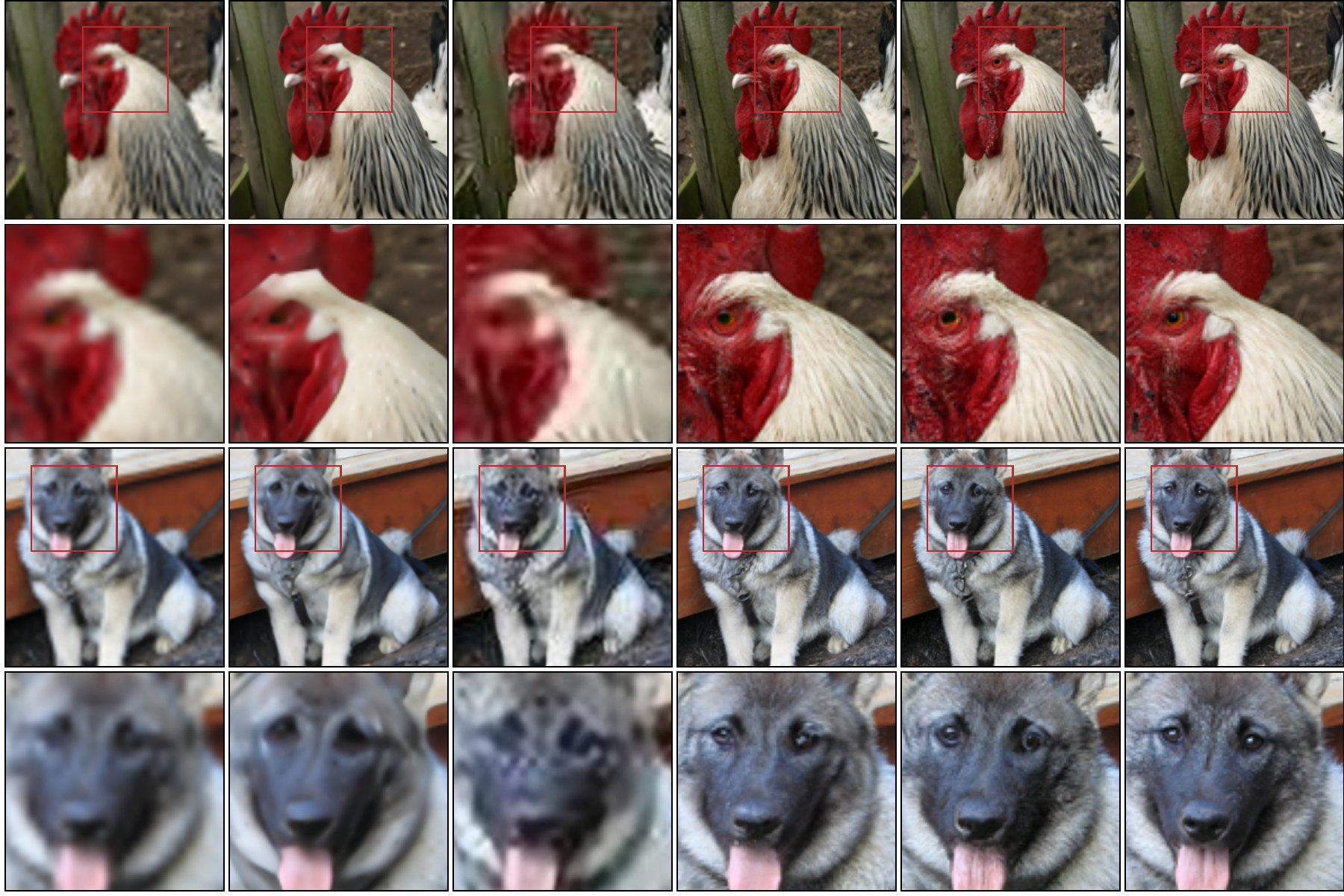}
        \put(4, 67){Bicubic}
        \put(21, 67){SRGAN~\cite{ledig2017photo}}
        \put(38, 67){LIIF~\cite{chen2021learning}}
        \put(55, 67){SR3~\cite{sr3}}
        \put(72, 67){PartDiff}
        \put(90, 67){HR}
    \end{overpic}
    \vspace{-2.5em}
    \caption{
        PartDiff and other \imsr{} methods on $64\times64\rightarrow256\times256$ image \sr
        applied to ImageNet test samples.
        Compared to others, PartDiff and SR3 generate highly realistic and visually pleasing textures.
        PartDiff achieves almost the same quality as SR3 with only half of the denoising steps.
        \ifdefined\withappendix
        More results can be found in the Appendix.
        \fi
    }
    \vspace{-0.8em}
    \label{fig:imagenet}
\end{figure*}

\subsection{Experiments on Natural Images}\label{sec:exp-natural-img}

\subsubsection{ImageNet}\label{sec:exp-imagenet}
We apply our method to $64\times64\rightarrow256\times256$ \sr{} on
the challenging ImageNet~\cite{deng2009imagenet} dataset with
over one million images of one thousand categories.
All models are trained with the training set and evaluated on the test set.
During training, we randomly crop a square image along the longer edge,
and test images are center cropped.
~\cref{fig:imagenet} compares PartDiff with other \sr{} methods including
GAN-based, Flow-based, and diffusion-based methods.
SRGAN~\cite{ledig2017photo} tends to generate blurry patterns with less detailed textures.
ESRGAN~\cite{wang2018esrgan}
improve the details to some extent but there is still an obvious gap with
real high-resolution images.
SR3 and PartDiff generate highly realistic textures that are close to real images.

We quantify the performance on ImageNet in two aspects:
1) consistency with low-resolution inputs and
2) reality of generated images.
To quantify the consistency, we compute the MSE between the downsampled outputs and low-resolution inputs.
The reality is measured by FID and inception score (IS).
~\cref{tab:imagenet-fid-is} summarize the quantitative comparisons on the ImageNet dataset.
\begin{table}[!htb]
    \centering
    \vspace{-1.2em}
    \caption{
        Quantitative results on ImageNet dataset~\cite{deng2009imagenet}.
    }\vspace{-1em}
    \newcommand{\CCa}{\cellcolor{gray!10}}
    \newcommand{\CCb}{\cellcolor{c2!10}}
    \resizebox{0.8\linewidth}{!}{
    \begin{tabular}[!htb]{lccc}
        \Xhline{2\arrayrulewidth}
        Model & Consist $\downarrow$ &  FID $\downarrow$ & IS $\uparrow$ \\
        \hline
        Reference & - & 1.9 & 240.8 \\
        Regression & 3.11 & 16.1 & 119.7 \\
        SRGAN & 27.7 & 13.9 & 162.5 \\
        \hline
        SR3 & 3.12 & 6.1 & 174.7 \\
        PartDiff ($K=25$) & 3.19 & 7.2 & 167.4 \\
        PartDiff ($K=50$) & 3.13 & 6.3 & 172.9 \\
        PartDiff ($K=75$) & 3.12 & 6.1 & 174.6 \\
        \hline
    \end{tabular}
    }
    \label{tab:imagenet-fid-is}
    \vspace{-1.5em}
\end{table}
As shown in~\cref{tab:imagenet-fid-is},
diffusion-based methods significantly outperform others in terms
of FID and IS scores, revealing that diffusion models generate highly realistic images.
All the methods achieve similar consistency with the low-resolution input,
but SR3 and PartDiff outperform others by a slight margin.
Our reimplemented SR3 performs slightly worse but still outperforms other competitors.
PartDiff achieves 157.4 in IS score and 11.2 in FID using only 25 denoising steps.
Using 50 denoising steps, PartDiff surpasses GAN-based methods by a significant margin.
To achieve similar performance, SR3 requires 100 denoising steps.

\subsubsection{Facial Images}
~\cref{fig:face128} and ~\cref{fig:face512} show some results
on $16\times16\rightarrow128\times128$ and $64\times64\rightarrow512\times512$ face \sr, respectively.
The upsampling scale is set to $8$ to show the generation ability of diffusion models.
As shown in~\cref{fig:face128,fig:face512}, regression-based approach produce
blurry output images without fine details.
We found that both implicit representation-based methods, i.e. LIIF~\cite{chen2021learning},
and GAN-based method, i.e. SRGAN~\cite{ledig2017photo}, generate visually unpleasing
results under this setting.
In contrast, diffusion-based methods generate visually pleasing images with highly-realistic details  such as hair
and eyes (see the zoomed-in patches).
PartDiff generates almost the same image quality as SR3 does but with fewer denoising steps.
\ifdefined\withappendix
More visualization results can be found in the Appendix.
\fi

Many previous works~\cite{chen2018fsrnet,dahl2017pixel,sr3} have pointed out that
PSNR and SSIM do not correlate well with human perception when the generated images
contain a large volume of high-frequency details.
We quantify the performance of facial \imsr{} in two aspects:
1) perceptual reality and
2) consistency to low-resolution input.
The Learned Perceptual Image Patch Similarity (LPIPS)~\cite{zhang2018perceptual} LPIPS can well judge the perceptual similarity between two images
and is used to measure perceptual reality.
We compute the MSE between the downsampled outputs and the low-resolution inputs
as the consistency metric.
\begin{table}[!htb]
    \centering
    \vspace{-1em}
    \caption{
        Quantitative results on facial image \sr.
    }
    \vspace{-1.2em}
    \newcommand{\CCa}{\cellcolor{gray!10}}
    \newcommand{\CCb}{\cellcolor{c2!5}}
    \resizebox{0.98\linewidth}{!}{
    \begin{tabular}[!htb]{lcc|cc}
        \Xhline{2\arrayrulewidth}
        & \multicolumn{2}{c|}{$16\times16\rightarrow128\times128$} & \multicolumn{2}{c}{$64\times64\rightarrow512\times512$} \\
        Model & Consist $\downarrow$ & LPIPS$\downarrow$ & Consist $\downarrow$ & LPIPS$\downarrow$  \\
        \hline
        Regression & 2.71 & 0.635 & 2.63 & 0.4981 \\
        SRGAN~\cite{ledig2017photo} & 127.1 & 0.210 & 131.8 & 0.2821 \\
        ESRGAN & 158.2 & 0.174 & 164.2 & 0.2795 \\
        \hline
        SR3 & 2.68 & 0.0799 & 2.61 &  0.2175 \\
        \CCb PartDiff ($K=25$) & \CCb 2.71 & \CCb 0.1491 & \CCb 2.69 & \CCb 0.2217 \\
        \CCb PartDiff ($K=50$) & \CCb 2.70 & \CCb 0.1299 & \CCb 2.64 & \CCb 0.2173 \\
        \hline
    \end{tabular}
    }
    \label{tab:face}
    \vspace{-0.2em}
\end{table}
As shown in~\cref{tab:face}, diffusion-based methods reduce the consistency error from hundreds
to less  than 3 on both $16\times16\rightarrow128\times128$ and $64\times64\rightarrow512\times512$ \sr,
and improve the LPIPS score as well.
With 25 denoising steps, partial diffusion achieves comparable performance with GANs.
Using 50 denoising steps, partial diffusion almost matches the performance of SR3
with 100 denoising steps.

\subsection{Comparison to SR3}\label{sec:comp-sr3}
In this experiment, we compare our method with SR3~\cite{sr3}.
As pointed out by many previous works~\cite{chen2018fsrnet,dahl2017pixel,sr3},
automated evaluation metrics, e.g., PSNR and SSIM, do not correlate
well with human perception when the generated images contain a large volume of
high-frequency details.
To capture the subtle differences between PartDiff and SR3,
we use human evaluation to compare the image quality.
For each dataset, we randomly select 500 test samples for human evaluation
because full-dataset evaluation on these datasets is time-consuming.
We adopt the two alternative force choice (2AFC) paradigm
where  human subjects are shown a low-resolution image and two shuffled \sr{}
results, and forced to select a result of `high-quality'.
Then we calculate the fool rate which is the percentage of PartDiff results being
selected.
The fool rate reflects how realistic PartDIff generated images are.
Theoretically, if PartDiff perfectly approximates SR3, 
the fool rate will be infinitely close to 50\%.
\begin{figure}[!htb]
    \vspace{-0.8em}
    \centering
    \begin{overpic}[width=0.95\linewidth]{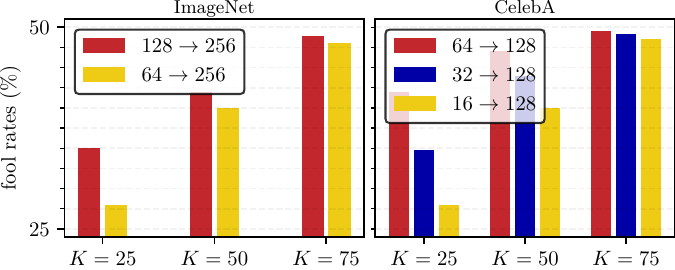}
    \end{overpic}
    \vspace{-1em}
    \caption{
        Fool rates with various $K$  on CelebA and ImageNet datasets.
    }
    \label{fig:fool-rate}
    \vspace{-0.5em}
\end{figure}
The fool rates in ~\cref{fig:fool-rate} consistently improve with the raising of denoising steps.
And the fool rate and the upsampling scale are inversely proportional:
under the same denoising steps, large upsampling factors lead to low fool rates.
This is not surprising because, as analyzed in~\cref{sec:diff-lr-hr},
the gap between the latent states of low and high-resolution images
become more significant with larger upsampling factors,
and such a gap harms the accuracy of using the low-resolution latent states
as the proxy for high-resolution latent states.
In general, the results in~\cref{fig:fool-rate} prove that PartDiff is able
to generate samples of almost  the same quality as SR3 with less
denoising steps.

\subsection{Application to Downstream Tasks}\label{sec:exp-downstream}
In this experiment, we apply \sr{} outputs to downstream tasks.
We test 3 tasks:
1) zonal segmentation on T2-weighted prostate MR images,
2) face recognition on LFW~\cite{huang2008labeled} dataset, and
3) image classification on ImageNet~\cite{deng2009imagenet} dataset.
Models for each task are pretrained with high-resolution images.
During testing, we first downsample the test images,
and then upsample them using different
\sr{} approaches.
The upsampled images are finally used as test images as the input to downstream
models.
Under this setting, models might achieve higher performance if it generates
samples that match the distribution of real images.

\subsubsection{Prostate Zonal Segmentation}\label{sec:zonal}
\begin{figure}
    \centering
    \vspace{1em}
    \begin{overpic}[width=0.9\linewidth]{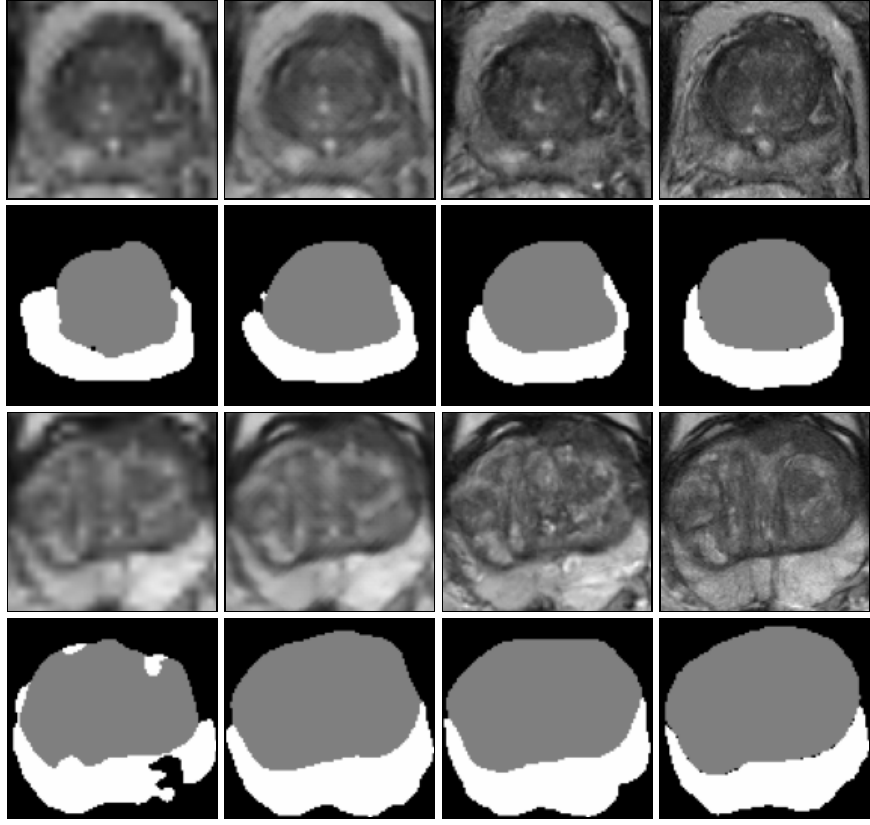}
        \put(6, 95.5){Bicubic}
        \put(26, 95.5){SRGAN~\cite{ledig2017photo}}
        \put(55, 95.5){PartDiff}
        \put(79, 95.5){HR (GT)}
    \end{overpic}
    \vspace{-1.5em}
    \caption{
        Example input images and output masks of zonal segmentation.
        The input images are upsampled by $\times4$ using different methods.
    }
    \label{fig:zonal-seg}
    \vspace{-1em}
\end{figure}
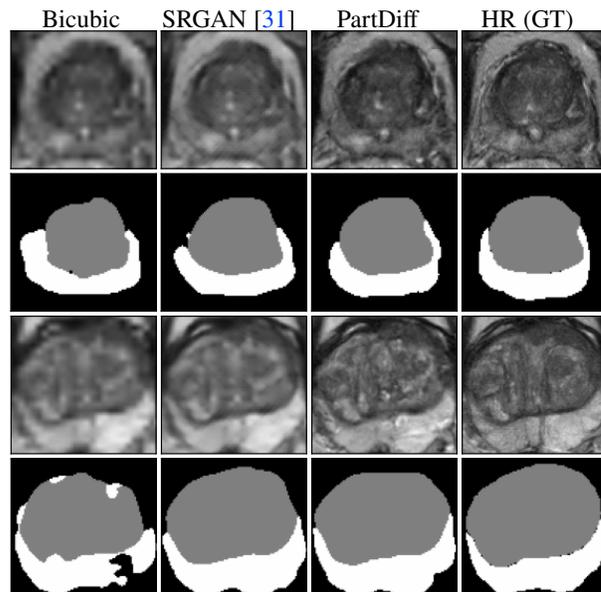

Prostate zonal segmentation is an important step in automatic prostate cancer detection,
and a suspicious lesion should be analyzed differently in different prostate zones,
due to variations in image appearance and cancer prevalence~\cite{israel2020multiparametric}.
In this experiment, we test the performance of zonal segmentation using images upsampled
with different methods.
We use a pretrained model~\cite{hung2022cat} which segments prostate into
peripheral zone (PZ) and transition zone (TZ).
\begin{figure}
    \centering
    \begin{overpic}[width=1\linewidth]{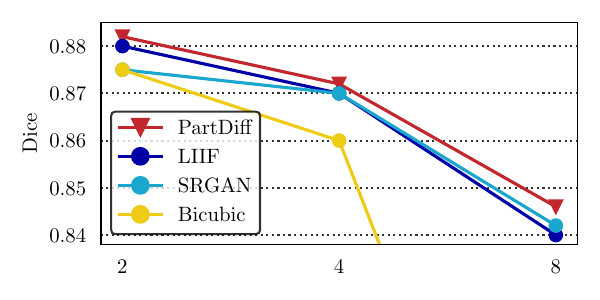}
    \end{overpic}
    \vspace{-3.5em}
    \caption{
        Zonal segmentation performance under different upsampling factors.
        The test images are downsampled and then upsampled by various \sr
        models.
    }
    \label{fig:zonal}
    \vspace{-0.5em}
\end{figure}
~\cref{fig:zonal} compares the dice coefficients of segmentation results
under various upsampling factors
and ~\cref{fig:zonal-seg} shows some zonal segmentation results
of images upsampled by $\times4$.

Our method consistently achieves higher segmentation performance
under various upsampling factors, and the performance gap is becoming
wider at large upsampling factors.
Interestingly, although our upsampled images are visually much more realistic
than SRGAN, the improvements in segmentation are not
as noticeable as the visual differences.
This reveals that visual realism is not well correlated with segmentation
quality.

\subsubsection{Face Recognition}
We use ArcFace~\cite{deng2019arcface} face recognition model as the baseline to evaluate
the performance of various \sr{} methods on face recognition.
The baseline model is pretrained on MS1MV2~\cite{guo2016ms} dataset and the performance
is evaluated on the LFW~\cite{huang2008labeled} dataset.
We use the same preprocessing protocol as ~\cite{deng2019arcface} to align and crop the test images.
The test images are cropped into $112\times 112$ patches.
These patches are first downsampled by a factor of 4 with bicubic interpolation,
and then upsample to the original resolution using different \sr{} approaches.
All the \sr{} models are trained on FFHQ dataset~\cite{karras2019style}.
\begin{table}[!htb]
    \centering
    \vspace{-0.5em}
    \caption{
        Super-resolution applied to face recognition on the LFW dataset.
        The test images are first bicubically downsampled ($\times4$) 
        and then upsampled by different \sr{} approaches.
    }
    \vspace{-1em}
    \resizebox{0.58\linewidth}{!}{
    \begin{tabular}{c|c}
        \Xhline{2\arrayrulewidth}
        Method & LFW Acc\\
        \Xhline{0.6\arrayrulewidth}
        ArcFace~\cite{deng2019arcface} (baseline) & 99.43 \\
        Bicubic & 88.72 \\
        Regression & 89.75 \\
        SRGAN~\cite{ledig2017photo}  & 89.71 \\
        \hline
        SR3~\cite{sr3} ($T=100$) & \textbf{91.01} \\
        PartDiff ($K=25$) & 89.88 \\
        PartDiff ($K=50$) & 90.85 \\
        PartDiff ($K=75$) & \textbf{91.01} \\
        \Xhline{2\arrayrulewidth}
    \end{tabular}
    }
    \label{tab:face-reco}
\end{table}
As shown in ~\cref{tab:face-reco},
the regression-based model performs slightly better than bicubic interpolation,
and SRGAN performs even worse than bicubic interpolation.
Diffusion-based methods surpass all other methods and reach the highest accuracy
of 91.01\%.
The proposed method achieves the accuracy of 89.97\% and 90.00\% with 25 and 50 denoising steps,
which is very close to the accuracy of SR3 with 100 steps.

\subsubsection{Image Classification}
We test the classification accuracy of \sr{} outputs
on the ImageNet dataset.
We use a pretrained ResNet-50 model as the baseline.
We perform $4\times$ \sr{} on the center-cropped $56\times56$ validation
images.
Classification accuracy on the ImageNet validation set is
summarized in~\cref{tab:img-cls}.
\begin{table}[!htb]
    \centering
    \vspace{-1em}
    \caption{
        Imagenet classification accuracy of a pretrained ResNet-50 network
        on the validation set.
        All test images are bicubically downsampled ($\times4$) and then upsampled
        by different models.
    }
    \vspace{-1em}
    \resizebox{0.70\linewidth}{!}{
    \begin{tabular}{lcc}
        \Xhline{2\arrayrulewidth}
        Method & Top-1 Acc & Top-5 Acc \\
        \hline
        Baseline & 0.748 & 0.920 \\
        Bicubic & 0.608  & 0.811 \\
        Regrerssion & 0.617  & 0.827 \\
        \hline
        SR3 ($T=100$) & \textbf{0.683} & \textbf{0.880} \\
        PartDiff ($K=25$) & 0.632 & 0.851\\
        PartDiff ($K=50$) & 0.679  &  \textbf{0.880} \\
        PartDiff ($K=75$) & \textbf{0.682} & \textbf{0.880} \\
        \Xhline{2\arrayrulewidth}
    \end{tabular}
    }
    \label{tab:img-cls}
\end{table}
Diffusion-based methods achieve the best top-1 and top-5 accuracies, suggesting
the outputs align with real images the best.
Our method, with three-quarters of the denoising steps, achieves identical accuracies.
Using only half denoising steps, our method achieves the same top-5 accuracy with and
very close top-1 accuracy to SR3.


\section{Conclusions}\label{sec:conclusion}
We introduce the Partial Diffusion Model, a novel diffusion-based model for accelerated image super-resolution.
Our method accelerates plain denoising diffusion models by skipping some of the diffusion steps.
We first observed that the latent states of a pair of low- and high-resolution images gradually converge and become indistinguishable.
Based on this observation, we propose approximating an intermediate state in denoising a high-resolution image
with a low-resolution image’s latent, which allows us to skip many denoising steps.
We further propose a latent alignment mechanism that gradually interpolates between the latent states of
diffusing low- and high-resolution images to compensate for the error caused by the approximation.
Experiments on both MRI and natural images demonstrate that the proposed method significantly reduces
the number of denoising steps without sacrificing the quality of generation.
One possible limitation is that it is only applicable to conditional generation tasks where a conditional input
is used for the approximation.
Possible future work includes exploring its application in other conditional generation tasks such as image
denoising and image inpainting.

\bibliographystyle{IEEEtran}
\bibliography{mr-sr.bib}

\begin{thebibliography}{10}
\providecommand{\url}[1]{#1}
\csname url@samestyle\endcsname
\providecommand{\newblock}{\relax}
\providecommand{\bibinfo}[2]{#2}
\providecommand{\BIBentrySTDinterwordspacing}{\spaceskip=0pt\relax}
\providecommand{\BIBentryALTinterwordstretchfactor}{4}
\providecommand{\BIBentryALTinterwordspacing}{\spaceskip=\fontdimen2\font plus
\BIBentryALTinterwordstretchfactor\fontdimen3\font minus
  \fontdimen4\font\relax}
\providecommand{\BIBforeignlanguage}[2]{{%
\expandafter\ifx\csname l@#1\endcsname\relax
\typeout{** WARNING: IEEEtran.bst: No hyphenation pattern has been}%
\typeout{** loaded for the language `#1'. Using the pattern for}%
\typeout{** the default language instead.}%
\else
\language=\csname l@#1\endcsname
\fi
#2}}
\providecommand{\BIBdecl}{\relax}
\BIBdecl

\bibitem{irani1991improving}
M.~Irani and S.~Peleg, ``Improving resolution by image registration,''
  \emph{CVGIP: Graphical models and image processing}, vol.~53, no.~3, pp.
  231--239, 1991.

\bibitem{dong2015image}
C.~Dong, C.~C. Loy, K.~He, and X.~Tang, ``Image super-resolution using deep
  convolutional networks,'' \emph{{IEEE Transactions on Pattern Recognition and
  Machine Intelligence}}, vol.~38, no.~2, pp. 295--307, 2015.

\bibitem{johnson2016perceptual}
J.~Johnson, A.~Alahi, and L.~Fei-Fei, ``Perceptual losses for real-time style
  transfer and super-resolution,'' in \emph{{European Conference on Computer
  vision}}.\hskip 1em plus 0.5em minus 0.4em\relax Springer, 2016, pp.
  694--711.

\bibitem{goodfellow2014generative}
I.~Goodfellow, J.~Pouget-Abadie, M.~Mirza, B.~Xu, D.~Warde-Farley, S.~Ozair,
  A.~Courville, and Y.~Bengio, ``Generative adversarial nets,'' in
  \emph{{Neural Information Processing Systems}}, Z.~Ghahramani, M.~Welling,
  C.~Cortes, N.~Lawrence, and K.~Weinberger, Eds., vol.~27.\hskip 1em plus
  0.5em minus 0.4em\relax Curran Associates, Inc., 2014.

\bibitem{kingma2013auto}
D.~P. Kingma and M.~Welling, ``Auto-encoding variational bayes,'' in
  \emph{{International Conference on Learning Representations}}, 2014.

\bibitem{karras2018progressive}
T.~Karras, T.~Aila, S.~Laine, and J.~Lehtinen, ``Progressive growing of {GAN}s
  for improved quality, stability, and variation,'' in \emph{{International
  Conference on Learning Representations}}, 2018.

\bibitem{brock2018large}
A.~Brock, J.~Donahue, and K.~Simonyan, ``Large scale {GAN} training for high
  fidelity natural image synthesis,'' in \emph{{International Conference on
  Learning Representations}}, 2019.

\bibitem{kalchbrenner2017video}
N.~Kalchbrenner, A.~Oord, K.~Simonyan, I.~Danihelka, O.~Vinyals, A.~Graves, and
  K.~Kavukcuoglu, ``Video pixel networks,'' in \emph{{International Conference
  on Machine Learning}}.\hskip 1em plus 0.5em minus 0.4em\relax PMLR, 2017, pp.
  1771--1779.

\bibitem{brooks2022generating}
T.~Brooks, J.~Hellsten, M.~Aittala, T.-C. Wang, T.~Aila, J.~Lehtinen, M.-Y.
  Liu, A.~A. Efros, and T.~Karras, ``Generating long videos of dynamic
  scenes,'' in \emph{{Neural Information Processing Systems}}, 2022.

\bibitem{vandenoord16_ssw}
A.~{van den Oord}, S.~Dieleman, H.~Zen, K.~Simonyan, O.~Vinyals, A.~Graves,
  N.~Kalchbrenner, A.~Senior, and K.~Kavukcuoglu, ``{WaveNet: A Generative
  Model for Raw Audio},'' in \emph{Proc. 9th ISCA Workshop on Speech Synthesis
  Workshop (SSW 9)}, 2016, p. 125.

\bibitem{prenger2019waveglow}
R.~Prenger, R.~Valle, and B.~Catanzaro, ``Waveglow: A flow-based generative
  network for speech synthesis,'' in \emph{ICASSP 2019-2019 IEEE International
  Conference on Acoustics, Speech and Signal Processing (ICASSP)}.\hskip 1em
  plus 0.5em minus 0.4em\relax IEEE, 2019, pp. 3617--3621.

\bibitem{chen2018fsrnet}
Y.~Chen, Y.~Tai, X.~Liu, C.~Shen, and J.~Yang, ``Fsrnet: End-to-end learning
  face super-resolution with facial priors,'' in \emph{{IEEE/CVF Conference on
  Computer Vision and Pattern Recognition}}, 2018, pp. 2492--2501.

\bibitem{dahl2017pixel}
R.~Dahl, M.~Norouzi, and J.~Shlens, ``Pixel recursive super resolution,'' in
  \emph{{IEEE/CVF International Conference on Computer Vision}}, 2017, pp.
  5439--5448.

\bibitem{ho2020denoising}
J.~Ho, A.~Jain, and P.~Abbeel, ``Denoising diffusion probabilistic models,''
  \emph{{Neural Information Processing Systems}}, vol.~33, pp. 6840--6851,
  2020.

\bibitem{sr3}
C.~Saharia, J.~Ho, W.~Chan, T.~Salimans, D.~J. Fleet, and M.~Norouzi, ``Image
  super-resolution via iterative refinement,'' \emph{{IEEE Transactions on
  Pattern Recognition and Machine Intelligence}}, pp. 1--14, 2022.

\bibitem{li2022srdiff}
H.~Li, Y.~Yang, M.~Chang, S.~Chen, H.~Feng, Z.~Xu, Q.~Li, and Y.~Chen,
  ``Srdiff: Single image super-resolution with diffusion probabilistic
  models,'' \emph{Neurocomputing}, vol. 479, pp. 47--59, 2022.

\bibitem{fattal2007image}
R.~Fattal, ``Image upsampling via imposed edge statistics,'' in \emph{ACM
  SIGGRAPH 2007 papers}, 2007, pp. 95--es.

\bibitem{sun2008image}
J.~Sun, Z.~Xu, and H.-Y. Shum, ``Image super-resolution using gradient profile
  prior,'' in \emph{{IEEE/CVF Conference on Computer Vision and Pattern
  Recognition}}.\hskip 1em plus 0.5em minus 0.4em\relax IEEE, 2008, pp. 1--8.

\bibitem{shan2008fast}
Q.~Shan, Z.~Li, J.~Jia, and C.-K. Tang, ``Fast image/video upsampling,''
  \emph{ACM Transactions on Graphics (TOG)}, vol.~27, no.~5, pp. 1--7, 2008.

\bibitem{kim2010single}
K.~I. Kim and Y.~Kwon, ``Single-image super-resolution using sparse regression
  and natural image prior,'' \emph{{IEEE Transactions on Pattern Recognition
  and Machine Intelligence}}, vol.~32, no.~6, pp. 1127--1133, 2010.

\bibitem{freeman2002example}
W.~T. Freeman, T.~R. Jones, and E.~C. Pasztor, ``Example-based
  super-resolution,'' \emph{IEEE Computer graphics and Applications}, vol.~22,
  no.~2, pp. 56--65, 2002.

\bibitem{chang2004super}
H.~Chang, D.-Y. Yeung, and Y.~Xiong, ``Super-resolution through neighbor
  embedding,'' in \emph{{IEEE/CVF Conference on Computer Vision and Pattern
  Recognition}}, vol.~1.\hskip 1em plus 0.5em minus 0.4em\relax IEEE, 2004, pp.
  I--I.

\bibitem{glasner2009super}
D.~Glasner, S.~Bagon, and M.~Irani, ``Super-resolution from a single image,''
  in \emph{{IEEE/CVF International Conference on Computer Vision}}.\hskip 1em
  plus 0.5em minus 0.4em\relax IEEE, 2009, pp. 349--356.

\bibitem{freedman2011image}
G.~Freedman and R.~Fattal, ``Image and video upscaling from local
  self-examples,'' \emph{ACM Transactions on Graphics (TOG)}, vol.~30, no.~2,
  pp. 1--11, 2011.

\bibitem{yang2013fast}
J.~Yang, Z.~Lin, and S.~Cohen, ``Fast image super-resolution based on in-place
  example regression,'' in \emph{{IEEE/CVF Conference on Computer Vision and
  Pattern Recognition}}, 2013, pp. 1059--1066.

\bibitem{wang2015deep}
Z.~Wang, D.~Liu, J.~Yang, W.~Han, and T.~Huang, ``Deep networks for image
  super-resolution with sparse prior,'' in \emph{{IEEE/CVF International
  Conference on Computer Vision}}, 2015, pp. 370--378.

\bibitem{shi2016real}
W.~Shi, J.~Caballero, F.~Husz{\'a}r, J.~Totz, A.~P. Aitken, R.~Bishop,
  D.~Rueckert, and Z.~Wang, ``Real-time single image and video super-resolution
  using an efficient sub-pixel convolutional neural network,'' in
  \emph{{IEEE/CVF Conference on Computer Vision and Pattern Recognition}},
  2016, pp. 1874--1883.

\bibitem{sajjadi2017enhancenet}
M.~S. Sajjadi, B.~Scholkopf, and M.~Hirsch, ``Enhancenet: Single image
  super-resolution through automated texture synthesis,'' in \emph{{IEEE/CVF
  International Conference on Computer Vision}}, 2017, pp. 4491--4500.

\bibitem{li2019feedback}
Z.~Li, J.~Yang, Z.~Liu, X.~Yang, G.~Jeon, and W.~Wu, ``Feedback network for
  image super-resolution,'' in \emph{{IEEE/CVF Conference on Computer Vision
  and Pattern Recognition}}, 2019, pp. 3867--3876.

\bibitem{chen2021learning}
Y.~Chen, S.~Liu, and X.~Wang, ``Learning continuous image representation with
  local implicit image function,'' in \emph{{IEEE/CVF Conference on Computer
  Vision and Pattern Recognition}}, 2021, pp. 8628--8638.

\bibitem{ledig2017photo}
C.~Ledig, L.~Theis, F.~Husz{\'a}r, J.~Caballero, A.~Cunningham, A.~Acosta,
  A.~Aitken, A.~Tejani, J.~Totz, Z.~Wang \emph{et~al.}, ``Photo-realistic
  single image super-resolution using a generative adversarial network,'' in
  \emph{{IEEE/CVF Conference on Computer Vision and Pattern Recognition}},
  2017, pp. 4681--4690.

\bibitem{arjovsky2017wasserstein}
M.~Arjovsky, S.~Chintala, and L.~Bottou, ``Wasserstein gan,'' 2017.

\bibitem{metz2017unrolled}
L.~Metz, B.~Poole, D.~Pfau, and J.~Sohl-Dickstein, ``Unrolled generative
  adversarial networks,'' in \emph{{International Conference on Learning
  Representations}}, 2017.

\bibitem{heusel2017gans}
M.~Heusel, H.~Ramsauer, T.~Unterthiner, B.~Nessler, and S.~Hochreiter, ``Gans
  trained by a two time-scale update rule converge to a local nash
  equilibrium,'' \emph{{Neural Information Processing Systems}}, vol.~30, 2017.

\bibitem{sohl2015deep}
J.~Sohl-Dickstein, E.~Weiss, N.~Maheswaranathan, and S.~Ganguli, ``Deep
  unsupervised learning using nonequilibrium thermodynamics,'' in
  \emph{{International Conference on Machine Learning}}.\hskip 1em plus 0.5em
  minus 0.4em\relax PMLR, 2015, pp. 2256--2265.

\bibitem{nichol2021improved}
A.~Q. Nichol and P.~Dhariwal, ``Improved denoising diffusion probabilistic
  models,'' in \emph{{International Conference on Machine Learning}}.\hskip 1em
  plus 0.5em minus 0.4em\relax PMLR, 2021, pp. 8162--8171.

\bibitem{ho2022video_a}
J.~Ho, T.~Salimans, A.~A. Gritsenko, W.~Chan, M.~Norouzi, and D.~J. Fleet,
  ``Video diffusion models,'' in \emph{{Neural Information Processing
  Systems}}, A.~H. Oh, A.~Agarwal, D.~Belgrave, and K.~Cho, Eds., 2022.

\bibitem{chen2021wavegrad}
N.~Chen, Y.~Zhang, H.~Zen, R.~J. Weiss, M.~Norouzi, and W.~Chan, ``Wavegrad:
  Estimating gradients for waveform generation,'' in \emph{{International
  Conference on Learning Representations}}, 2021.

\bibitem{li2022diffusionlm}
X.~L. Li, J.~Thickstun, I.~Gulrajani, P.~Liang, and T.~Hashimoto,
  ``Diffusion-{LM} improves controllable text generation,'' in \emph{{Neural
  Information Processing Systems}}, A.~H. Oh, A.~Agarwal, D.~Belgrave, and
  K.~Cho, Eds., 2022.

\bibitem{lu2022dpm}
C.~Lu, Y.~Zhou, F.~Bao, J.~Chen, C.~Li, and J.~Zhu, ``Dpm-solver: A fast ode
  solver for diffusion probabilistic model sampling in around 10 steps,'' in
  \emph{{Neural Information Processing Systems}}, 2022.

\bibitem{salimans2022progressive}
T.~Salimans and J.~Ho, ``Progressive distillation for fast sampling of
  diffusion models,'' in \emph{{International Conference on Learning
  Representations}}, 2022.

\bibitem{zheng2023truncated}
H.~Zheng, P.~He, W.~Chen, and M.~Zhou, ``Truncated diffusion probabilistic
  models and diffusion-based adversarial auto-encoders,'' in
  \emph{{International Conference on Learning Representations}}, 2023.

\bibitem{song2019generative}
Y.~Song and S.~Ermon, ``Generative modeling by estimating gradients of the data
  distribution,'' \emph{{Neural Information Processing Systems}}, vol.~32,
  2019.

\bibitem{hyvarinen2005estimation}
A.~Hyv{\"a}rinen and P.~Dayan, ``Estimation of non-normalized statistical
  models by score matching.'' \emph{{Journal of Machine Learning Research}},
  vol.~6, no.~4, 2005.

\bibitem{vincent2011connection}
P.~Vincent, ``A connection between score matching and denoising autoencoders,''
  \emph{Neural computation}, vol.~23, no.~7, pp. 1661--1674, 2011.

\bibitem{song2020improved}
Y.~Song and S.~Ermon, ``Improved techniques for training score-based generative
  models,'' \emph{{Neural Information Processing Systems}}, vol.~33, pp.
  12\,438--12\,448, 2020.

\bibitem{paszke2019pytorch}
A.~Paszke, S.~Gross, F.~Massa, A.~Lerer, J.~Bradbury, G.~Chanan, T.~Killeen,
  Z.~Lin, N.~Gimelshein, L.~Antiga \emph{et~al.}, ``Pytorch: An imperative
  style, high-performance deep learning library,'' \emph{{Neural Information
  Processing Systems}}, vol.~32, 2019.

\bibitem{loshchilov2018decoupled}
I.~Loshchilov and F.~Hutter, ``Decoupled weight decay regularization,'' in
  \emph{{International Conference on Learning Representations}}, 2019.

\bibitem{karras2019style}
T.~Karras, S.~Laine, and T.~Aila, ``A style-based generator architecture for
  generative adversarial networks,'' in \emph{{IEEE/CVF Conference on Computer
  Vision and Pattern Recognition}}, 2019, pp. 4401--4410.

\bibitem{deng2009imagenet}
J.~Deng, W.~Dong, R.~Socher, L.-J. Li, K.~Li, and L.~Fei-Fei, ``Imagenet: A
  large-scale hierarchical image database,'' in \emph{{IEEE/CVF Conference on
  Computer Vision and Pattern Recognition}}.\hskip 1em plus 0.5em minus
  0.4em\relax Ieee, 2009, pp. 248--255.

\bibitem{clark2013cancer}
K.~Clark, B.~Vendt, K.~Smith, J.~Freymann, J.~Kirby, P.~Koppel, S.~Moore,
  S.~Phillips, D.~Maffitt, M.~Pringle \emph{et~al.}, ``The cancer imaging
  archive (tcia): maintaining and operating a public information repository,''
  \emph{Journal of digital imaging}, vol.~26, pp. 1045--1057, 2013.

\bibitem{Turkbey2019PIRADS}
T.~B, R.~AB, and H.~M. et~al, ``Prostate imaging reporting and data system
  version 2.1: 2019 update of prostate imaging reporting and data system
  version 2,'' \emph{Eur. Urol.}, vol.~76, no.~3, pp. 340--351, 2019.

\bibitem{sood2019anisotropic}
R.~Sood and M.~Rusu, ``Anisotropic super resolution in prostate mri using super
  resolution generative adversarial networks,'' in \emph{International
  Symposium on Biomedical Imaging}.\hskip 1em plus 0.5em minus 0.4em\relax
  IEEE, 2019, pp. 1688--1691.

\bibitem{wang2018esrgan}
X.~Wang, K.~Yu, S.~Wu, J.~Gu, Y.~Liu, C.~Dong, Y.~Qiao, and C.~Change~Loy,
  ``Esrgan: Enhanced super-resolution generative adversarial networks,'' in
  \emph{{European Conference on Computer vision Workshops}}, 2018, pp. 0--0.

\bibitem{zhang2018perceptual}
R.~Zhang, P.~Isola, A.~A. Efros, E.~Shechtman, and O.~Wang, ``The unreasonable
  effectiveness of deep features as a perceptual metric,'' in \emph{CVPR},
  2018.

\bibitem{huang2008labeled}
G.~B. Huang, M.~Mattar, T.~Berg, and E.~Learned-Miller, ``Labeled faces in the
  wild: A database forstudying face recognition in unconstrained
  environments,'' in \emph{Workshop on faces in'Real-Life'Images: detection,
  alignment, and recognition}, 2008.

\bibitem{israel2020multiparametric}
B.~Israel, M.~van~der Leest, M.~Sedelaar, A.~R. Padhani, P.~Zamecnik, and J.~O.
  Barentsz, ``Multiparametric magnetic resonance imaging for the detection of
  clinically significant prostate cancer: what urologists need to know. part 2:
  interpretation,'' \emph{European urology}, vol.~77, no.~4, pp. 469--480,
  2020.

\bibitem{hung2022cat}
A.~L.~Y. Hung, H.~Zheng, Q.~Miao, S.~S. Raman, D.~Terzopoulos, and K.~Sung,
  ``Cat-net: A cross-slice attention transformer model for prostate zonal
  segmentation in mri,'' \emph{IEEE TMI}, vol.~42, no.~1, pp. 291--303, 2022.

\bibitem{deng2019arcface}
J.~Deng, J.~Guo, N.~Xue, and S.~Zafeiriou, ``Arcface: Additive angular margin
  loss for deep face recognition,'' in \emph{{IEEE/CVF Conference on Computer
  Vision and Pattern Recognition}}, 2019, pp. 4690--4699.

\bibitem{guo2016ms}
Y.~Guo, L.~Zhang, Y.~Hu, X.~He, and J.~Gao, ``Ms-celeb-1m: A dataset and
  benchmark for large-scale face recognition,'' in \emph{{European Conference
  on Computer vision}}.\hskip 1em plus 0.5em minus 0.4em\relax Springer, 2016,
  pp. 87--102.

\end{thebibliography}

\newcommand{\AddPhoto}[1]{\includegraphics%
[width=1in,height=1.25in,clip,keepaspectratio]{figures/photos/#1}}

\ifdefined\withauthorbio
\begin{IEEEbiography}[\AddPhoto{kai.jpg}]{Kai Zhao}
    received his Ph.D. degree in Computer Science from Nankai University in 2020,
    and his M.S. and B.S. in electrical engineering from Shanghai University in 2014 and 2017, respectively.
    He is currently a postdoctoral researcher at the University of California,
    Los Angeles.
    Prior to joining UCLA, he was a research scientist at Tencent.
    His research interests include medical image analysis,
    computer vision, and statistical learning,
    with a specific focus on low-data deep learning,
    where the amount of annotated data is limited.
    More information can be found on his homepage \url{https://kaizhao.net}.
\end{IEEEbiography}
\vspace{-.5in}

\begin{IEEEbiography}[\AddPhoto{alex.pdf}]{Alex Ling Yu Hung}
    is currently a Ph.D. student in Computer Science at the University of California, Los Angeles (UCLA), advised by Prof. Kyunghyun Sung and Prof. Demetri Terzopoulos. 
    He received his Master's degree in Biomedical Engineering from Carnegie Mellon University in 2021, and Bachelor's degree from Peking University in 2019 while majoring Biomedical Engineering and minoring in Computer Science. 
    His main research focus is on medical image analysis, generative models, and general computer vision. 
    More information can be found on his homepage \url{https://web.cs.ucla.edu/~alexhung/}.
\end{IEEEbiography}
\vspace{-.5in}

\begin{IEEEbiography}[\AddPhoto{kaifeng.jpg}]{Kaifeng Pang}
    is currently a Master student in the 
    Electrical and Computer Engineering Department of the 
    University of California, Los Angeles (UCLA), 
    working with Dr Kai Zhao and Prof. Kyunghyun Sung. 
    He received his Bachelor's degree from Nanjing University in 2022. 
    His research interests include medical image analysis and deep learning.
\end{IEEEbiography}
\vspace{-.5in}

\begin{IEEEbiography}[\AddPhoto{haoxin}]{Haoxin Zheng}
    is currently a Ph.D. student in the
    Computer Science Department of the University of California, Los Angeles (UCLA),
    advised by Prof. Kyunghyun Sung and Prof. Fabien Scalzo.
    He received his Master's degree from UCLA in 2019,
    and Bachelor's degree from Nankai University in 2017.
    His research interests include medical image analysis, 
    multi-modality learning, and computer vision.
\end{IEEEbiography}
\vspace{-.5in}

\begin{IEEEbiography}[\AddPhoto{kyung}]{Kyunghyun Sung}
    received the Ph.D. degree
    in electrical engineering from the University of
    Southern California, Los Angeles, in 2008. From
    2008 to 2012, he finished his postdoctoral training
    with the Departments of Radiology, Stanford. In
    2012, he joined the Department of Radiological
    Sciences, University of California at Los Angeles,
    Los Angeles (UCLA). He is currently an Associate
    Professor of Radiology, where his research primarily 
    focuses on the development of novel medical
    imaging methods and artificial intelligence using magnetic resonance
    imaging (MRI). In particular, his research group  \url{https://mrrl.ucla.edu/sunglab/}
    is currently focused on developing advanced deep learning algorithms and
    quantitative MRI techniques for early diagnosis, treatment guidance, and
    therapeutic response assessment for oncologic applications. Such developments
    can offer more robust and reproducible measures of biologic markers
    associated with human cancers.
\end{IEEEbiography}
\fi

\clearpage
\pagebreak

\ifdefined\withappendix
\begin{figure*}[ht]
    \vspace{1.5em}
    \begin{overpic}[width=1\linewidth]{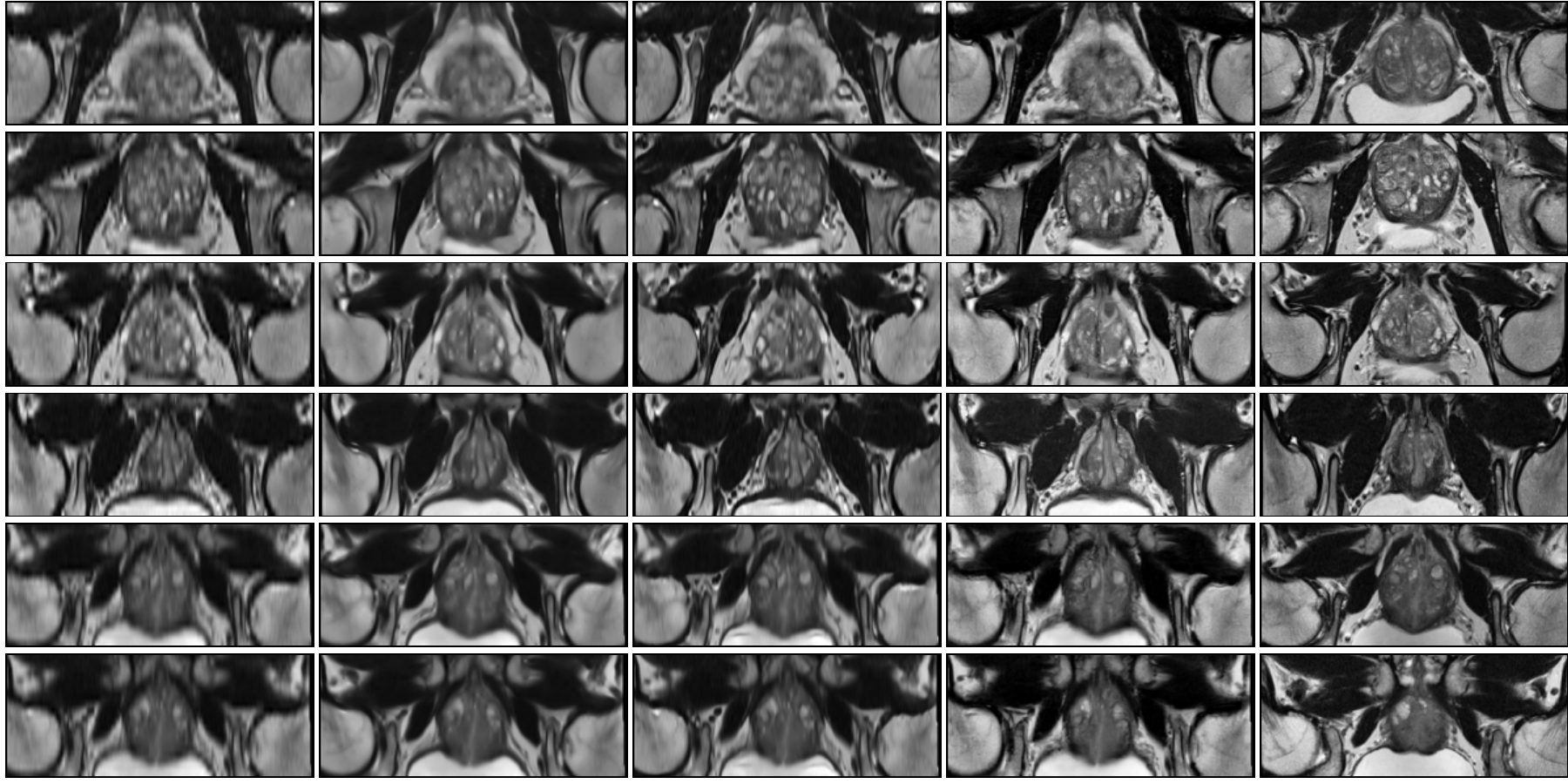}
        \put(9, 52.4){Input }
        \put(5,50.7){($d\times w$), axial}
        \put(24, 51){SRGAN~\cite{ledig2017photo,sood2019anisotropic} }
        \put(46, 51){LIIF~\cite{chen2021learning}}
        \put(66, 51){PartDiff}
        \put(83, 52.4){Visual reference}
        \put(83, 50.7){($h\times w$), coronal}
    \end{overpic}
    \caption{
        Additional results on through-plane T2-weighted prostate MR image super-resolution.
    }\label{fig:through-plane-more}
\end{figure*}

\begin{figure*}[!hb]
    \vspace{1.5em}
    \begin{overpic}[width=1\linewidth]{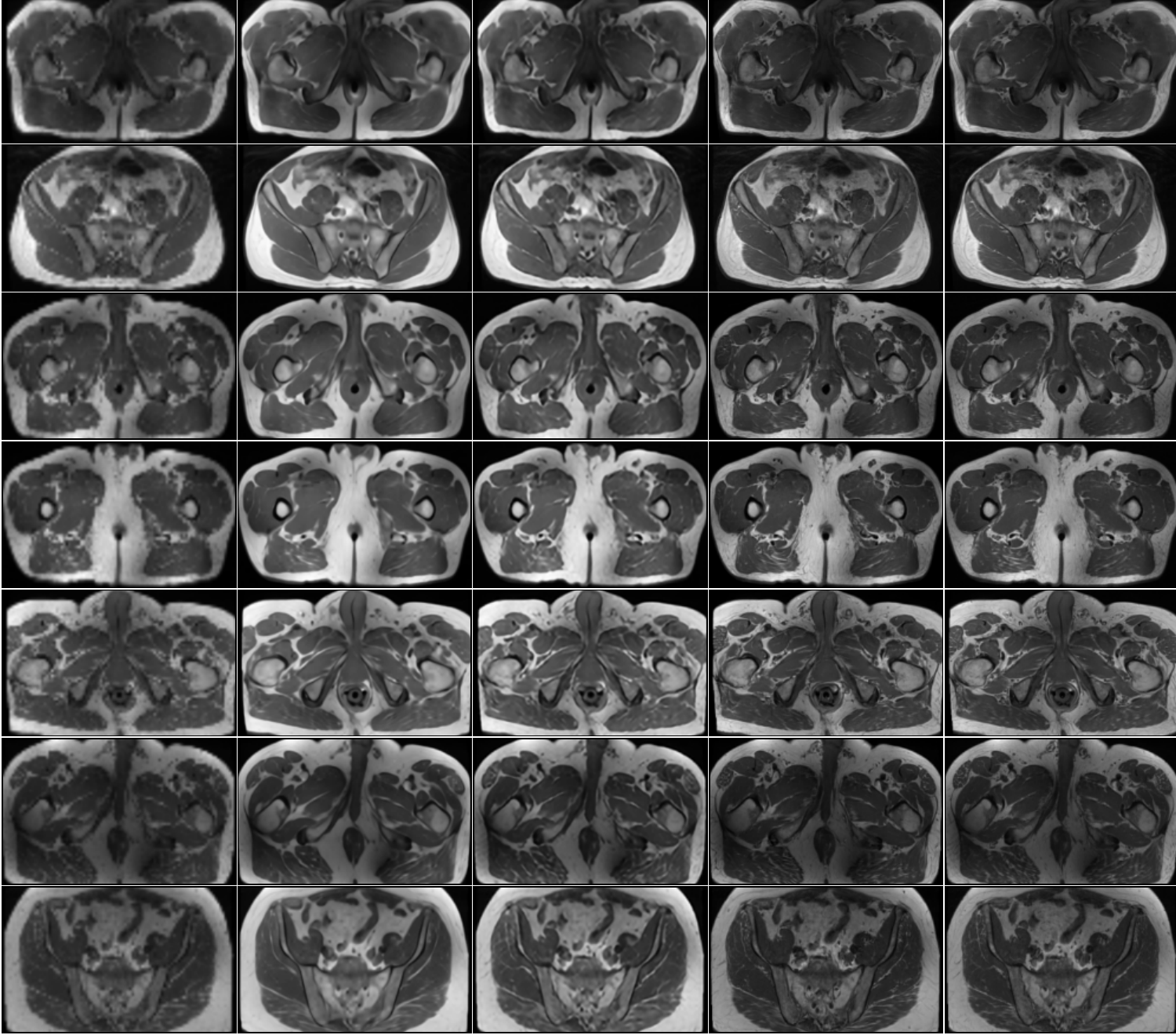}
        \put(5,88.8){Bicubic}
        \put(22,88.8){Regression}
        \put(43,88.8){SR3~\cite{sr3}}
        \put(63,88.8){PartDiff}
        \put(82,88.8){Reference}
    \end{overpic}
    \caption{
        Additional results  on T1-weighted prostate MR \imsr{} ($80\times80\rightarrow 320\times320$).
    }\label{fig:t1w-results-more}
\end{figure*}

\begin{figure*}[!hb]
    \vspace{1.5em}
    \begin{overpic}[width=1\linewidth]{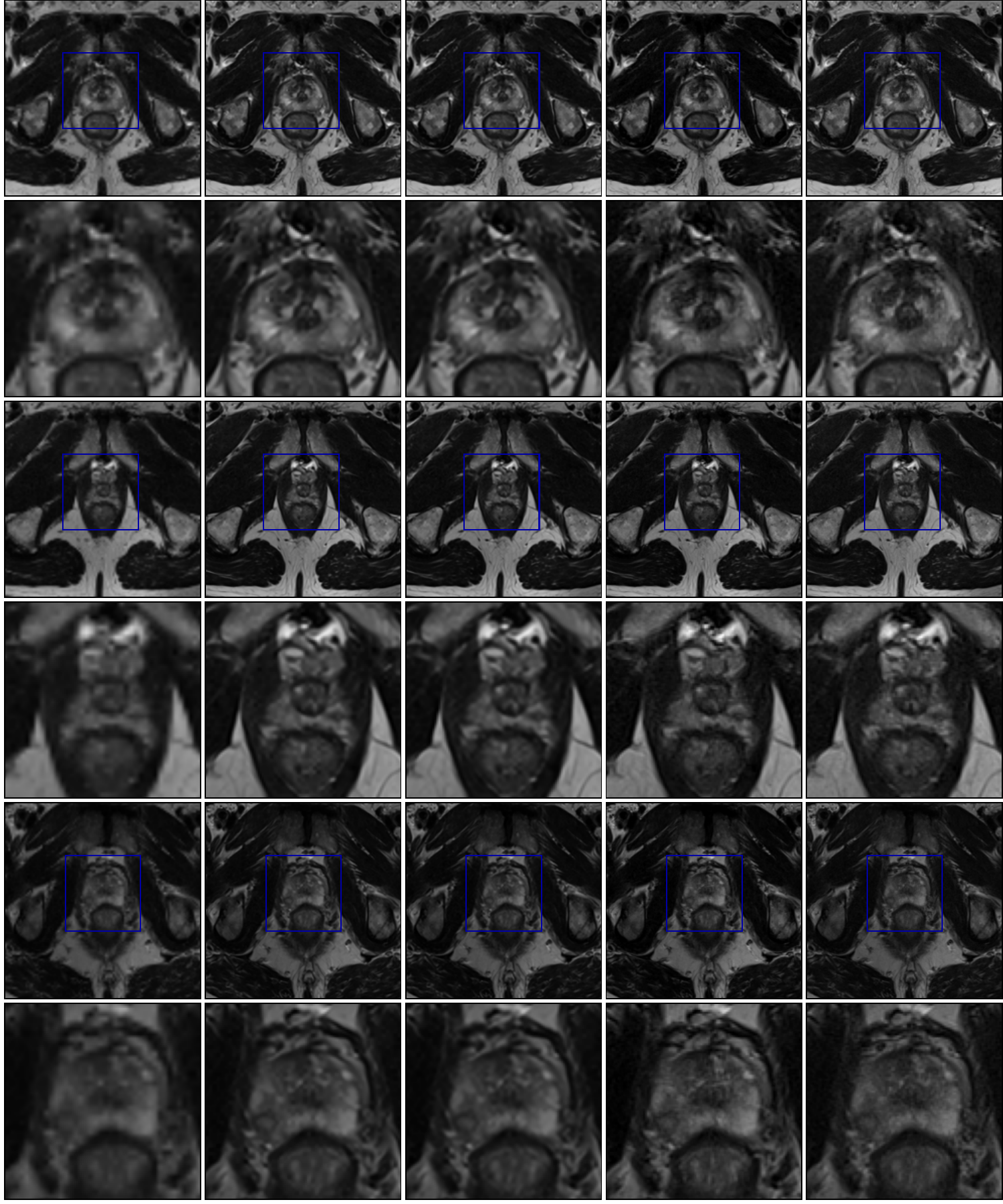}
        \put(5,101){Bicubic}
        \put(22,101){Regression}
        \put(37,101){SR3~\cite{sr3}}
        \put(55,101){PartDiff}
        \put(72,101){Reference}
    \end{overpic}
    \caption{
        Additional results on T2-weighted prostate MR \imsr{} ($160\times160\rightarrow 320\times320$).
    }\label{fig:t2w-results-more}
\end{figure*}

\begin{figure*}[!hb]
    \vspace{1.5em}
    \begin{overpic}[width=1\linewidth]{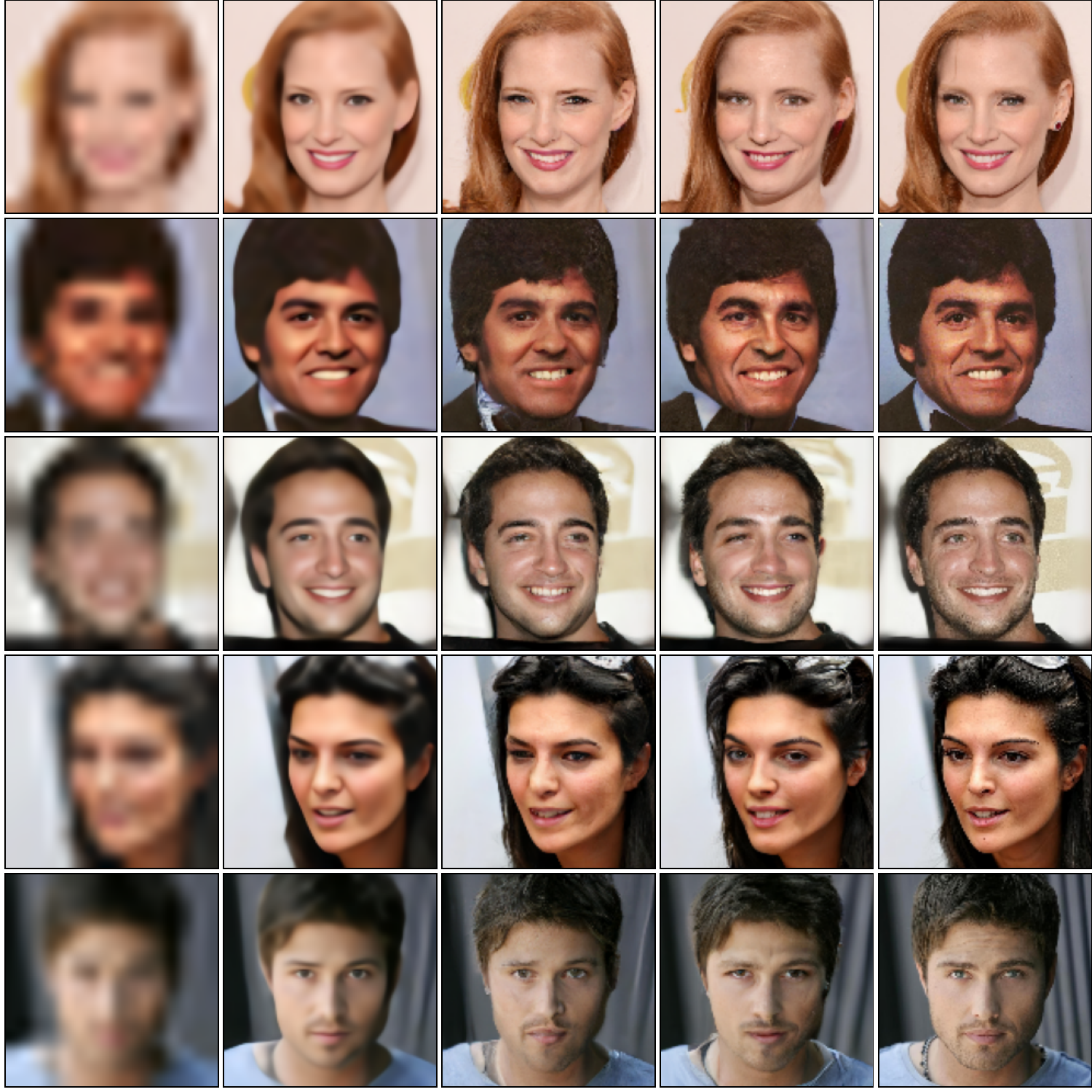}
        \put(7,101){Bicubic}
        \put(25,101){Regression}
        \put(45,101){SR3~\cite{sr3}}
        \put(65,101){PartDiff}
        \put(86,101){Reference}
    \end{overpic}
    \caption{
        Additional results ($16\times16\rightarrow 128\times128$) on the CelebA dataset.
    }\label{fig:face128-more}
\end{figure*}

\begin{figure*}
    \vspace{1.5em}
    \begin{overpic}[width=1\linewidth]{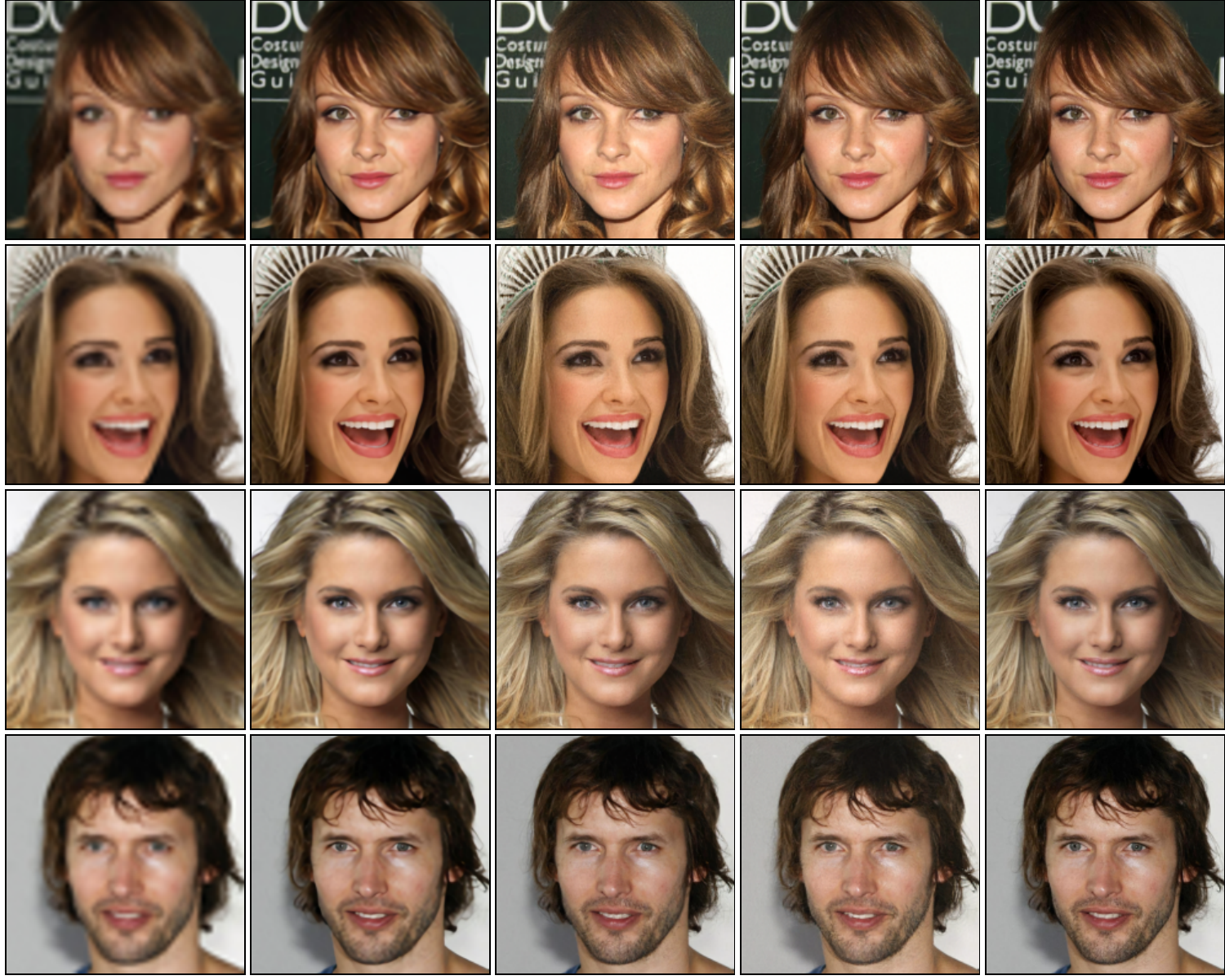}
        \put(7,81){Bicubic}
        \put(26,81){Regression}
        \put(46,81){SR3~\cite{sr3}}
        \put(66,81){PartDiff}
        \put(86,81){Reference}
    \end{overpic}
    \caption{
        Additional results ($64\times64\rightarrow 512\times512$) on the CelebA dataset.
        The model is trained on FFHQ~\cite{karras2019style} dataset.
    }\label{fig:face512-more}
\end{figure*}

\fi

\end{document}